\documentclass[sigconf]{acmart}
\AtBeginDocument{%
  \providecommand\BibTeX{{%
    \normalfont B\kern-0.5em{\scshape i\kern-0.25em b}\kern-0.8em\TeX}}}

\copyrightyear{2024}
\acmYear{2024}
\setcopyright{rightsretained}
\acmConference[CHI '24]{Proceedings of the CHI Conference on Human Factors in Computing Systems}{May 11--16, 2024}{Honolulu, HI, USA}
\acmBooktitle{Proceedings of the CHI Conference on Human Factors in Computing Systems (CHI '24), May 11--16, 2024, Honolulu, HI, USA}
\acmDOI{10.1145/3613904.3642875}
\acmISBN{979-8-4007-0330-0/24/05}

\usepackage[utf8]{inputenc}
\usepackage{subcaption}
\usepackage{multirow}
\usepackage{enumitem}
\usepackage[utf8]{inputenc}
\begin{document}

\title[Accessible Public Robots]{Co-designing Accessible Public Robots: Insights from People with Mobility Disabilities, Robotic Practitioners, and Their Collaborations}

\author{Howard Ziyu Han}
\email{ziyuh@andrew.cmu.edu}
\orcid{0009-0008-5556-7297}
\affiliation{%
  \institution{Human-Computer Interaction Institute, Carnegie Mellon University}
  \streetaddress{5000 Forbes Ave}
  \city{Pittsburgh}
  \state{Pennsylvania}
  \country{USA}
  \postcode{15213}
}

\author{Franklin Mingzhe Li}
\email{mingzhe2@cs.cmu.edu}
\orcid{0000-0003-4995-4545}
\affiliation{%
  \institution{Human-Computer Interaction Institute, Carnegie Mellon University}
  \streetaddress{5000 Forbes Ave}
  \city{Pittsburgh}
  \state{Pennsylvania}
  \country{USA}
  \postcode{15213}
}

\author{Alesandra Baca-Vazquez}

\email{alesandrabaca@utexas.edu}
\orcid{0009-0000-9936-9710}
\affiliation{%
  \department{School of Information}
  \institution{University of Texas at Austin }
  \streetaddress{1616 Guadalupe St Suite #5.202}
  \city{Austin}
  \state{Texas}
  \country{USA}
  \postcode{15213}
}

\author{Daragh Byrne}
\orcid{0000-0001-7193-006X}
\email{daragh@cmu.edu}
\affiliation{%
    \department{School of Architecture}
    \institution{Carnegie Mellon University }
  \streetaddress{5000 Forbes Ave}
  \city{Pittsburgh}
  \state{Pennsylvania}
  \country{USA}
  \postcode{15213}
}

\author{Nikolas Martelaro}
\authornote{Co-senior authors contributed equally to this research.}
\email{nikmart@cmu.edu}
\orcid{0000-0002-1824-0243}
\affiliation{%
  \institution{Human-Computer Interaction Institute, Carnegie Mellon University}
  \streetaddress{5000 Forbes Ave}
  \city{Pittsburgh}
  \state{Pennsylvania}
  \country{USA}
  \postcode{15213}
}

\author{Sarah E Fox}
\authornotemark[1]
\email{sarahfox@cmu.edu}
\orcid{0000-0002-7888-2598}
\affiliation{%
  \institution{Human-Computer Interaction Institute, Carnegie Mellon University}
  \streetaddress{5000 Forbes Ave}
  \city{Pittsburgh}
  \state{Pennsylvania}
  \country{USA}
  \postcode{15213}
}
\renewcommand{\shortauthors}{Han et al.}

\begin{abstract}
% 150 words
Sidewalk robots are increasingly common across the globe. Yet, their operation on public paths poses challenges for people with mobility disabilities (PwMD) who face barriers to accessibility, such as insufficient curb cuts. We interviewed 15 PwMD to understand how they perceive sidewalk robots. Findings indicated that PwMD feel they have to compete for space on the sidewalk when robots are introduced. We next interviewed eight robotics practitioners to learn about their attitudes towards accessibility. Practitioners described how issues often stem from robotic companies addressing accessibility only after problems arise. Both interview groups underscored the importance of integrating accessibility from the outset. Building on this finding, we held four co-design workshops with PwMD and practitioners in pairs. These convenings brought to bear accessibility needs around robots operating in public spaces and in the public interest. Our study aims to set the stage for a more inclusive future around public service robots.

\end{abstract}





 
%%
%% This command processes the author affiliation and title
\maketitle

\section{Introduction}
Alongside the widening use of service robots generally, sidewalk robots are among the newest innovations to be introduced into public spaces~\cite{rosenthal2020forgotten,bennett2021accessibility,weinberg2023sharing}. 
Several companies such as FedEx, Starship, and Uber Eats are currently testing or deploying robots designed to deliver food, medicine, and other small cargo~\cite{Stars8578372:online}. Although there are few public deployments of robots on sidewalks, some states have proactively prepared legislation to allow more widespread use of the technology; Pennsylvania, for instance, recently defined delivery robots as ``pedestrians'' under state law, ostensibly giving them the same legal rights as human residents. Further, regulations allow the robots to travel at speeds up to 12 miles per hour and weigh 500 pounds unloaded~\cite{BillI4329789:online}.
However, when sidewalk robots navigate pedestrian walkways, most people they interact with will not be users of the robot, but are instead non-users who ``happen to be there'' ~\cite{rosenthal2020forgotten}. 
Among those passersby will be people with mobility disabilities (PwMD)---such as wheelchair users---whose needs should be considered in any debate or development of such devices that operate in public~\cite{bennett2021accessibility}.
Yet, ongoing studies have revealed conflicts between PwMD and sidewalk robots~ \cite{bennett2021accessibility,weinberg2023sharing,bhat2022confused,ackerman2019lessons,han2023robot}. Drawing on interviews with multiple stakeholders, including disability advocates, company representatives, and city government officials, Bennett et al.~\cite{bennett2021accessibility} found that sidewalk robots can re-introduce access barriers long fought against by the disability rights community~\cite{meyers2002barriers}. 
Additionally, the second author on~\cite{bennett2021accessibility}, a wheelchair user, experienced a dangerous encounter with a sidewalk robot that stopped on a curb cut, blocking her safe passage to the other side of the road. 
After sharing a Tweet recounting the experience~\cite{ackerman2019lessons}, the company removed the robot from operating in public for a brief time. 
However, she received backlash for ``\textit{whining}'', given that she ``\textit{was not hit by a car''.} 
This incident---and subsequent responses---highlights a clear knowledge gap around PwMD's interactions with sidewalk robots and the need to maintain accessible public walkways.

Given these prior incidents and the potential for harm, the needs of PwMD are critical to consider when designing robots that travel on public sidewalks. 
However, aside from a limited number of research studies~\cite{a2014wayfinding}, accessible sidewalk robot interactions are often under-considered. 
To understand why, it is necessary to recognize the ways design decisions shape how robots are (or are not) embedded with accessibility features and are later integrated into society. 
For example, how practitioners acknowledge and incorporate PwMD's needs throughout the design and deployment process dictates the safety and societal implications of these robots. While much prior work in accessible Human-Robot Interaction (HRI) has engaged roboticists in co-design activities with people with disabilities (PwD)~\cite{valenciaCodesigningSociallyAssistive2021,fernaeusWhereThirdWave2009,leeStepsParticipatoryDesign2017,Azenkot2016robotguide}, few works reflect on the organizational practices that guide how robotics practitioners value, evaluate, and ensure accessibility when developing robots for public space. 

\begin{figure*}[h]
    \centering
    \includegraphics[width=\textwidth]{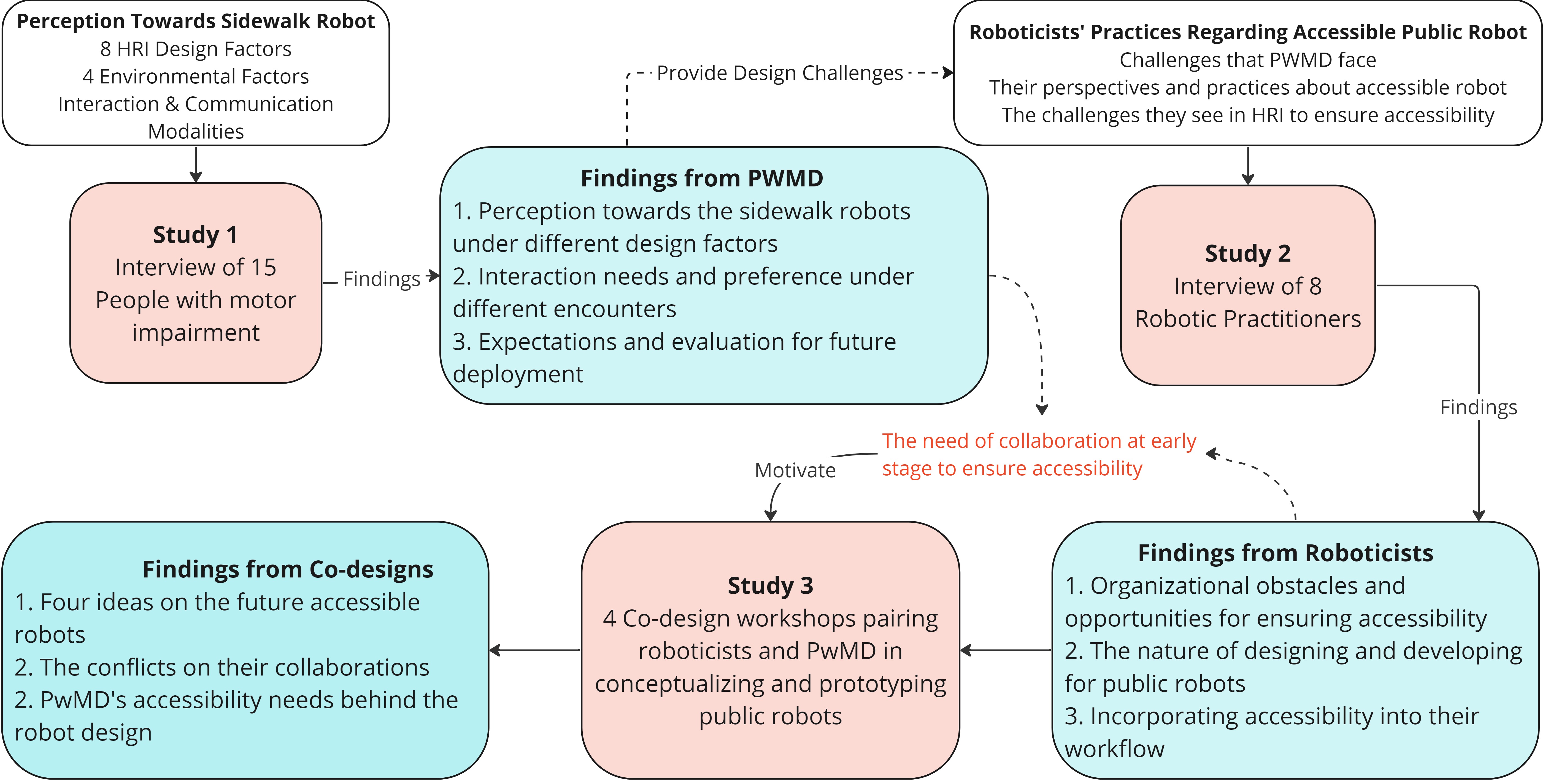}
    \caption{This flowchart illustrates the progressive approach taken in our three-part study and its key outcomes. The insights gained from the first study, which involved interviewing PwMD, were integrated into the interview framework conducted with robotic practitioners in study 2. The collective findings from both studies then formed the basis for the collaborative workshops in study 3.}
    \label{fig:paper-structure}
    \Description{This figure outlines the three-phase research process into the accessible design of sidewalk robots. Phase one involved interviewing 15 individuals with motor impairments to understand their perceptions and expectations based on various HRI design and environmental factors. In phase two, 8 robotic practitioners were interviewed to address the design challenges identified in phase one and explore organizational approaches to accessibility. The insights from the first two phases underscored the need for early collaboration, paving the way for phase three, which consisted of co-design workshops where roboticists and individuals with motor impairments collaborated to conceptualize accessible sidewalk robots.}
\end{figure*}

This paper aims to understand the complex relationship between sidewalk robots and PwMD in order to inform the design and operation of these robots in a safe and socially responsible way. To this end, we pose the following research questions: 

\begin{itemize}

\item \textbf{RQ1:} How do PwMD perceive the presence of robots on public walkways? How do they imagine encountering and interacting with these robots, given different HRI design factors and different environmental constraints?

\item \textbf{RQ2:} How do roboticists in industry and academia view the challenges raised by PwMD? What are the obstacles to and opportunities for improving their current practices in order to design future public robots that are more accessible? 

\item \textbf{RQ3:} How might robotics practitioners and PwMD conceptualize and prototype public robots together? How might the ideas and accessibility needs, in turn, inform current HRI practice?

\end{itemize}
To address these questions, we interviewed 15 PwMD to gain a deeper understanding of their perceptions related to sidewalk robot designs (RQ1). This exploration illuminated the potential impacts of robots on PwMDs navigation experiences. We further interviewed eight roboticists from both industrial and academic backgrounds. These interviews focused on identifying the organizational challenges and opportunities pivotal to enhancing accessibility in public robot design (RQ2). Building on findings from each set of interviews, we organized four co-design workshops, inviting PwMD and practitioners, working in pairs, to collaboratively conceptualize an accessible public service robot (RQ3). This process is summarized in Fig.\ref{fig:paper-structure}. 

Our study offers three key contributions to the HCI community: First, it provides detailed qualitative insights into PwMD's perspectives on sidewalk robots. Second, this work reflects on the contemporary dynamics of accessible robots through the lens of robotics industry practitioners. Finally, through co-design workshops, participants showcase accessible robot design ideas and offer perspectives on integrating accessibility into future public robot design.

\section{Related Work}
This work is situated at the intersection of sidewalk accessibility and HRI studies. Section\ref{r1-inaccessible} draws on scholarship examining the current  
challenges that PwMD have navigating public spaces and new mobility technologies for addressing some of these issues. 
Section\ref{sec:r2-public robots} reviews ethnographic observations of and experiments with sidewalk robots in public spaces, while Section\ref{sec:r3-robots impacts} focuses on the influence of public robots on sidewalk accessibility. 
Finally, Section\ref{r4-HRIdesign} reviews accessible HRI design guidelines and participatory processes that inform our study, focusing on design factors essential to public robots.  

\subsection{Inaccessible Public Space and Mobility Technology}  
\label{r1-inaccessible}
In the United States, regulations require that sidewalks be accessible to all individuals, including those using mobility devices such as wheelchairs \cite{PublicAccommodationsCommercial}. Unfortunately, for people with disabilities (PWD), using sidewalks can present challenges that impact their quality of life \cite{ferreiraProposalSidewalkAccessibility2007,kirchnerDesignedDeterCommunity2008}, independence \cite{mitchellPedestrianMobilitySafety2006}, physical activity levels \cite{mitchellPedestrianMobilitySafety2006}, and ability to participate fully in society \cite{ferreiraProposalSidewalkAccessibility2007}. These challenges are often due to barriers created by poor sidewalk design or construction, insufficient maintenance, or natural terrain features \cite{ferreiraProposalSidewalkAccessibility2007}. 
Studies on sidewalk accessibility have identified the challenges experienced by people with disabilities on a daily basis. Gharebaghi and Amin categorized these barriers into spatial (e.g., surface quality, slopes) and social (e.g., transport policy) factors \cite{gharebaghiIntegrationSocialEnvironment2018}. Mehmet et al.~identified various physical barriers, including uneven surfaces, obstructions, and insufficient width that impair sidewalk accessibility for PwMD \cite{mehmetAccessibilityDisabledPeople2009}. 

Work by Froehlich and colleagues uses design and technology to address sidewalk accessibility challenges  \cite{froehlichFutureUrbanAccessibility2022c}. For instance, their team's research has explored the potential of crowdsourcing platforms and deep learning techniques to map sidewalk conditions, allowing for the assessment of sidewalk navigability \cite{sahaProjectSidewalkWebbased2019}.
Teams such as Kasemsuppakorn and Karimi have used similar data \cite{kasemsuppakorn2008data} to develop wheelchair routing systems that can avoid obstacles and impassable sidewalks \cite{kasemsuppakorn2009personalised,kasemsuppakorn2015understanding}.
While such technology is helping PwMD better navigate sidewalks with fixed issues, the introduction of robots to public spaces raises questions about how PwMD navigation strategies may be affected.
Our research focuses on how PwMD will interact with robots while also managing other fixed sidewalk issues.
Work exploring prior deployments of public robots may suggest some ways in which they will alter public spaces and people's interaction with them on the sidewalk.

\subsection{Robots in Public}
\label{sec:r2-public robots}
% The deployment of new mobility technology can affect both spatial and social factors regarding the accessibility of the sidewalk. 
Currently, there are various deployments of robots in public spaces, both in research contexts and as part of new services.
Real-world deployments of robots in public include security robots that monitor parking lots and pedestrian malls\footnote{\url{https://www.knightscope.com}} and dog-like robots to patrol public spaces and encourage social distancing during the COVID-19 pandemic \cite{Toh_2020} or support police operations for dirty, dangerous, and dull tasks \cite{Bushwick_2021}.
The most widely deployed robots in public are delivery robots deployed by companies such as Starship, Kiwibot, Amazon, and FedEx.
These robots often travel slowly along public sidewalks to deliver goods to people.
In general, these robots do not often aim to interact with people, except when passing.
However, prior work by HRI research teams suggests that even when robots have incidental interaction with non-users in public, they will nonetheless cause changes in how people behave, thus making it important for robot development teams to understand how the public might interact with their robots \cite{rosenthal2020forgotten}.

Prior work in public HRI studies suggests some ways that people will interact with different robots in public.
Yang et al. \cite{yang2015trashbot} used Wizard-of-Oz field studies to develop socially acceptable movement and interactions for a robotic trashcan, finding that many people would actively ignore the robot if they did not want it to approach, would speak to the robot to summon it or request it leave, and would try to entice the robot with trash.
A more recent deployment in New York City has replicated these results and shown that people are generally receptive to trash barrel robots when they are in public squares \cite{buTrashBarrelRobots2023a}.
Work by Schneider et al. \cite{Schneider2022Ignoring} testing a humanoid robot in a Japanese mall also found that roughly half of the people walking by ignored the robot unless the robot spoke to people and asked to not be ignored. 
Thunberg and Ziemke \cite{thunbergArePeopleReady2020} found that many people in a European train station were not accustomed to interacting with a humanoid robot (Pepper by Aldebaran).
\cite{dobrosovestnova2022little} found that some people in Estonia helped delivery robots stuck in the snow, though they raise ethical issues about people providing free labor for delivery companies by helping the robots.
\cite{weinberg2023sharing} also observed people helping or ignoring stuck robots in a US city.
These results suggest a split among people's desire to interact with robots in public.
Moreover, the works specifically observing delivery robots suggest that the robots have similar challenges on the sidewalk as PwMDs and that they too may become obstacles impeding walkways.

\subsection{Public robot's impacts on sidewalk accessibility}
\label{sec:r3-robots impacts}
We now consider prior work that further explores the potential impacts robots may have on the accessibility of pedestrian walkways. 
Weinberg et al. \cite{weinberg2023sharing} found that robots could cause distractions and obstructions with different sidewalk users, such as blocking part of a sidewalk when stuck or causing dogs to become perturbed, by barking or lunging at the robots, and causing a minor scene on the sidewalk.
% Sara et al. \cite{nielsenUsingUserGeneratedYouTube2022} collected user-generated videos to explore unguided interactions between people and service robots in public spaces, identifying various circumstances in which these interactions can break down, such as when robots fail to detect individuals who want to interact with them or when the robots interrupt other people's interactions.
Gehrke and colleagues \cite{gehrkeEvaluationSidewalkAutonomous} found that delivery robots deployed on a college campus provided convenient last-mile delivery but also caused conflicts in walking paths and potentially unsafe travel conditions, especially for cyclists. 

Speaking directly to people with disabilities, Bennett et al.~\cite{bennett2021accessibility} highlight the safety concerns and real-world incidents of people with motor-related disabilities when encountering delivery robots on sidewalks, such as blocking access to a curb cut and preventing a wheelchair user to get from the street to back to the sidewalk. 
Bhat and Zhao \cite{bhat2022confused} explored the experiences of individuals with visual impairments who directly interact with different mobile service robots, describing these robots as ``dangerous'' and ``unfamiliar moving obstacles'' due to their lack of predictable behavior or clear intent.
Due to such incidents and the potential harm that they may cause pedestrians and PwMDs, researchers such as Salvini et al.~\cite{salviniSafetyConcernsEmerging2022} have recommended government regulation of robot deployments in public areas to address the psychological and physical safety risks that arise when pedestrians interact with sidewalk robots in highly crowded street environments. 
Thomasen~\cite{thomasenRobotsRegulationChanging2020}, however, argues that regulations must be carefully considered as different strategies for managing robots in public may help some people while encroaching on the rights of other's access to public space.

Our research builds upon these prior works exploring the accessibility issues of public robots to understand the potential challenges that PwMDs believe could exist.
Further, we also explore the perspectives of roboticists in considering accessibility during robot design.
While regulation is one proposed way for maintaining accessibility, it will also need to be paired with accessible robot design strategies that roboticists can build upon, as described in the next section.

\subsection{Accessible design strategies in Human-robot interaction}
\label{r4-HRIdesign}
Both HRI and HCI communities have proposed design frameworks for robot designers to follow~\cite{bartneckDesigncentredFrameworkSocial2004,chita-tegmarkAssistiveRobotsSocial2021}. 
However, there is less existing work on designing accessible service robots. 
We found only one proposal, evaluated by Qbilat \cite{qbilatProposalAccessibilityGuidelines2021a}, that specifically addressed accessibility in social robots by incorporating six other accessibility guidelines based on computer interface design (e.g. WCAG 2.0~\cite{initiativewaiWCAGOverview}, IBM accessibility~\cite{Acces2711621:online}). 
However, accessibility recommendations based on screen-based interfaces may not provide useful guidance for a robot's physical form and movement behavior.
Prior HRI studies have shown various factors of robot behavior design that may affect people's perception of safety and trust in robots. 
Previous experiments have shown that robot size, speed, and approaching behavior can alter people's trust in robots in indoor environments \cite{mutlu2008robots}. 
Further work has explored proxemics and people's comfort with robots at different distances and when approaching from different directions~\cite{takayama2009influences,walters2009empirical,samarakoon2022review}. 
Overall, there are many factors robot designers must consider in making robot movements safe and accessible.

Due to the complexity of creating safe and accessible robot interaction and the current lack of clear guidelines and understanding of people's responses to different robot designs, many teams approach designing robot behaviors through co-design workshops.
Prior works on assistive robotics have found success in co-developing robots tailored to the needs of people with disabilities ~\cite{valenciaCodesigningSociallyAssistive2021,fernaeusWhereThirdWave2009,leeStepsParticipatoryDesign2017}.
% These studies have leveraged their early collaborations to complement practitioners' knowledge and also shared insights on improving the accessibility of robots. However, most of the work focused more on assistive robots, rather than public robots. 
Based on these prior successes, researchers exploring public robots are also exploring participatory design approaches.
Early work by \cite{Azenkot2016robotguide} engaged people with vision disabilities (both designers and non-experts) through interviews, group co-design sessions, and Wizard-of-oz testing sessions to design robots that provide navigation guidance.
Tian et al. \cite{tianUserExpectationsRobots2020b} used participatory prototyping methods that involved programming robot interactions with the support of an expert and viewing them in a simulation to imagine how a humanoid robot could interact with people in public spaces.
Sumartojo et al. \cite{sumartojo2021imagining} used images of public spaces overlaid with different types of robots to elicit reactions from people about where such robots would or would not be appropriate.

Our research seeks to build upon the aforementioned participatory design research on public robots to better understand how the design choices involved in developing and deploying sidewalk robots might influence existing accessibility challenges.
While Participatory Design methods have been shown to be fruitful in generating new robot concepts and surfacing important concerns from users, there is little work engaging PwMDs in discussing robots on public sidewalks.
Our work builds on the methods of prior works by leveraging interviews and co-design sessions involving imagery and simulations \cite{Azenkot2016robotguide,tianUserExpectationsRobots2020b} to consider how robots on sidewalks should interact with PwMDs, described in Section\ref{sec:methods}.
Our study seeks to bring together insights from PwMD and roboticists on the design and presence of sidewalk robots. 
By learning from both groups and facilitating a collaborative design process between PwMDs and roboticists, we aim to inform accessible robot development in public space---whose intricacies haven't been fully examined.

% By presenting an array of existing design solutions and factors, we aimed to offer a comprehensive overview of current robot forms and behaviors, thereby encouraging participants to articulate their perceptions in relation to existing designs.
% These workshops aimed to explore the potential of a more inclusive design process and offered a platform for these two groups to communicate and ideate together. We explored how their collaborative work could inform and enrich the design of public space robots.

\section{Methods}
\label{sec:methods}
We adopted a multi-phased approach to examine the complex interplay between sidewalk robots, their design, and the perceptions they engender among PwMD and robotic practitioners. To understand how PwMD perceive sidewalk robots and investigate potential interaction methods, we interviewed 15 PwMD (refer to Table \ref{PwMD-table}) with an average age of 32.5 years old (SD=9.2). We also engaged with eight robotic practitioners (refer to Table \ref{Practitioner-table}), with an average age of 28.1 years old (SD=3.02). We recruited four practitioners from industry and four from academia, to understand the current design and development processes of sidewalk robots, as well as their thoughts on public robots more broadly. The recruitment criteria for roboticists required that they have at least two years of experience working on HRI design, research, or development. Lastly, we held four co-design workshops pairing PwMD and robotic practitioners. We invited both four PwMD and two roboticists from previous interviewees and recruited two new roboticists as needed based on the same criteria. 
The pairs were formed based mainly on their availability and people with overlapping available times were paired together. All the activities were held over video conference. Interviews lasted 60–80 minutes and co-design workshops lasted 90 minutes. Participants were compensated with a \$20 Amazon gift card for each study. Our university's Institutional Review Board (IRB) approved the recruitment and study procedure. 

\begin{figure*}[t]
    \centering
    \includegraphics[width=1.0\textwidth]{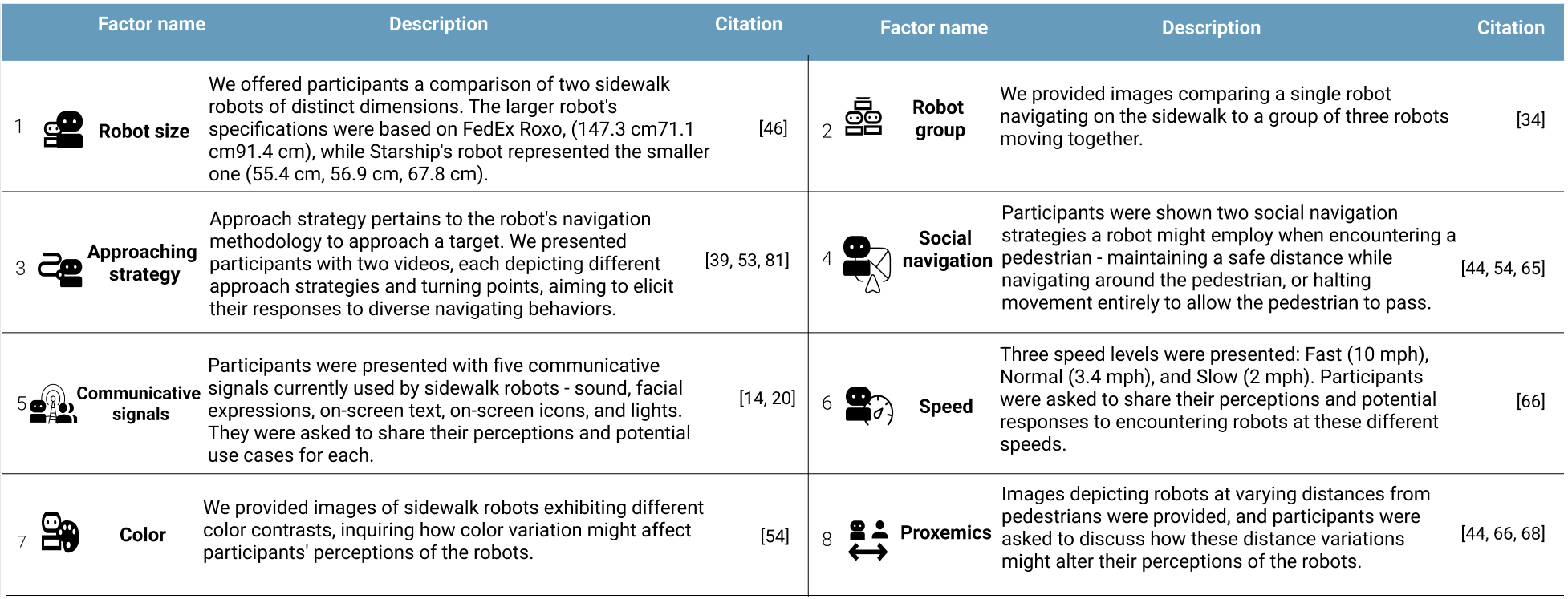}
    \caption{Drawing from existing literature, we identified eight critical HRI design factors. This figure provides an overview of these factors, elucidates how they were presented to the participants, and references the corresponding literature.}
    \label{fig:HRI_design_factor}
    \Description{This figure presents the eight human-robot interaction (HRI) design factors examined in the study, each represented with an icon and a brief description, along with a reference to the corresponding literature. The factors include 'Sizes,' illustrating the difference in dimensions between two robots; 'Approach Strategy,' showcasing various robotic navigation strategies; 'Communicative Signals,' depicting five communication modes utilized by sidewalk robots; 'Color,' exploring the influence of color contrasts on robot perception; 'Robot Group,' comparing reactions to a single robot versus a group; 'Social Navigation,' demonstrating different pedestrian encounter strategies; 'Speed,' indicating three potential speed levels for the robots; and 'Proxemics,' exploring the impact of different robot-pedestrian distances on perception.}
\end{figure*}

\begin{table*}[h]
\centering
\caption{Demographic information of PwMD}
\label{PwMD-table}
\begin{tabular}{@{}c l c l l@{}}
        \toprule
        Number & Gender & Age & Mobility Disability & Details \\
        \midrule
        P1 & Female & 25 & T5 complete Spinal Cord Injury & Use manual chairs \\
        P2 & Female & 52 & Idiopathic Brown Sequard Syndrome & Difficulty walking, not in a wheelchair anymore \\
        P3 & Male & 34 & L1 paraplegic & Wheelchair user, two years post-injury \\
        P4 & Male & 25 & T5 complete Spinal Cord Injury & Use manual wheelchairs \\
        P5 & Male & 36 & Quadriplegic & Uses a power wheelchair \\
        P6 & Male & 36 & L1 paraplegic & Manual wheelchair \\
        P7 & Female & 21 & spastic diplegic CP & Uses two loft-strand canes daily \\
        P8 & Female & 29 & C-5/6 quadriplegic & Use manual wheelchairs \\
        P9 & Male & 19 & physical disability from TK2D & Uses a power wheelchair \\
        P10 & Female & 35 & spinal cord injury & Walking difficulties \\
        P11 & Female & 40 & C5-C6 quadriplegic & Powerchair, paralyzed from armpits down \\
        P12 & Male & 20 & spastic diplegia cerebral palsy & Manual wheelchair \\
        P13 & Male & 36 & Quadriplegia from a SCI & Power wheelchair \\
        P14 & Male & 41 & Spinal cord injury & Power wheelchair \\
        P15 & Male & 39 & Quadriplegic & Manual wheelchair \\
        \bottomrule
    \end{tabular}
    \Description{A table labeled 'Demographic Information of PwMD' provides details on fifteen individuals with various mobility disabilities (PwMD). The columns from left to right are 'Number', representing each person uniquely with identifiers from P1 to P15, 'Gender' indicating male or female, 'Age' showing ages ranging from 19 to 52 years, 'Mobility Disability' giving details on the specific disability, and 'Details' offering further specifics about each individual’s mobility aids or condition. Disabilities noted include different levels of spinal cord injuries, quadriplegia, and spastic diplegic CP among others. The details column mentions the usage of different aids like manual and power wheelchairs, and loft-strand canes.}
\end{table*}

\begin{table*}[t]
\caption{Demographic Information of Robotic Practitioners}
\label{Practitioner-table}
\begin{tabular}{@{}p{0.8cm} p{0.8cm} p{0.5cm} p{2.8cm} p{2.5cm} p{6cm}@{}}
        \toprule
        Number & Gender & Age & Organization & Role & Description \\
        \midrule
        R1 & Male & 29 & U.S University & HRI researcher & Ph.D. students in Human-robot co-learning research \\
        R2 & Female & 26 & U.S University & HRI researcher & Ph.D. students researching the conversational interface and accessibility in human-robot interaction \\
        R3 & Female & 25 & U.S University & HRI researcher & Ph.D. students researching social navigation in human-robot interaction \\
        R4 & Male & 29 & Start-up Robot Company & Product Manager & PM at a room cleaning robot company with HRI knowledge; observed PDD for 6 months \\
        R5 & Female & 26 & Robot Department at a large company & Backend Engineer & Responsible for the backend of home robots; has experience working with visually-impaired groups \\
        R6 & Male & 34 & Robot Department at a large company & Safety Engineer & Has 7 years of experience in the safety team of a large company’s robot product \\
        R7 & Female & N$\backslash$A & Start-up Robot Company & User Experience Designer Manager & Founding member of the UX design team at an autonomous robot company with one year of experience \\
        R8 & Female & 28 & Canadian University & HRI researcher & Ph.D. students developing virtual simulations for people with disabilities interacting with autonomous vehicles \\
        \bottomrule
    \end{tabular}
    \Description{This table displays the demographic information of eight robotic practitioners labeled from R1 to R8. The data is structured into five columns which represent the number, gender, organization, role, and a description of each practitioner's role and expertise. Practitioners are associated with organizations such as U.S Universities, start-up robot companies, and large corporate robot departments. Their roles include various positions such as Human-Robot Interaction (HRI) researchers, product managers, backend engineers, safety engineers, and user experience designer managers. The description column provides a more detailed insight into their specific responsibilities and experience in their respective roles, with many engaged in Ph.D. research and others holding positions in companies where they work on aspects such as product safety and user experience.}
\end{table*}

\begin{table*}
    \centering
    \caption{Demographic Information of Workshops}
    \label{Workshop-table}
    \begin{tabular}{@{}c  l l@{}}
\toprule
Number & PwMD Participants & Roboticists Participants \\
\midrule
W1 & A 21 years old female cane user & A 26 years old HRI researcher on the conversational interface and accessibility \\
W2 & A 38 years old power female chair user & A 38 years old founder of a robotics start-up \\
W3 & A 40 years old male power chair user & A 30 years old HRI researcher on assistive technology \\
W4 & A 34 years old male power chair user & A 26 years old back-end robotic engineer \\
\hline
\end{tabular}
\Description{A table labeled 'Demographic Information of Workshops' presenting details of four workshops (W1 to W4). Each workshop lists a PwMD participant along with their age and mobility aid, and a roboticist participant along with their professional background. The participants' ages range from 21 to 40 years, and professions include HRI researcher and startup founder.}
\end{table*}

\subsection{Interviews with People with Mobility Disabilities}
In the interview with PwMD, we first asked our participants to share demographic information about themselves, such as age, gender, and their mobility impairment (should they wish to disclose). Then, we delved into the following topics: 

\subsubsection{Perception of Sidewalk Robot Design Factors}
To ensure participants were familiar with the concept of sidewalk robots, we showed a video of a functioning delivery robot and images of various delivery robots \cite{li2022freedom}. As a starting point to discuss design forms, we presented eight widely recognized design factors such as speeds and social navigation strategies (refer to \ref{Fig:socialnavigation}) from the HRI literature to each participant (refer to Fig\ref{fig:HRI_design_factor}~\cite{qbilat2021proposal,joosse2021making,takayama2009influences,story2022speed,mutlu2008robots,wojciechowska2019collocated,de2019pedestrian,bhimasta2019causes,mizumaru2019stop,sisbot2006mobile,lauckner2014hey,lee2009snackbot,fereday2006demonstrating,haring2014would}). We also selected four environmental barriers that might influence their perceptions about interacting with the robots including sidewalk width, slope, congestion, and weather~\cite{meyers2002barriers}. Subsequently, we invited our participants to discuss how these factors might influence their perceptions and navigation experiences.

\subsubsection{Interaction Modality}
We then delved into interaction modalities, sharing six explanatory images as reference --- voice interaction, gestures, apps, physical buttons, touchpads, and joysticks on the wheelchairs 
% (refer to Fig\ref{})
. Interviewees were encouraged to imagine how they might prefer to interact with the robots in different scenarios (e.g. if the robot is stuck in the road). 

\subsubsection{Future deployments}
The final portion of the interview was centered on discussions of future deployments including the functionalities interviewees expected sidewalk robots to have, and how they wanted sidewalk robots to be introduced or deployed. 

% This section of the interview sought to uncover the participants' needs in relation to potential interactions with the robots, and issues of usability that could emerge with different interaction modalities. 

\subsection{Interviews with Robotic Practitioners}
\label{methods:PwMD}
In the second set of interviews with practitioners, we centered our conversations around two primary questions: (1) How do robotic practitioners consider accessibility within their current practice, and what opportunities do they see for improvement? (2) Given the challenges highlighted by PwMD, how can public robots be made more accessible?

Initially, we delved into their perspectives on the state of accessible robots in their respective contexts, whether in companies or academic institutions. We sought to understand the obstacles they might face in ensuring accessibility for the robots they work on. We then explored potential avenues for change, posing questions like, "What values, methods, and research-based strategies are essential to incorporate accessibility features into public robots?" These questions were tailored to their specific roles (e.g., designer, researcher, or developer). Furthermore, we addressed the practical challenges outlined by PwMD in our initial interviews. For instance, we discussed issues like competition for sidewalk resources and invited suggestions on potential solutions.

\subsection{Co-Design Workshop}
\label{methods:Co-design}
Building on our findings from interviews, in the final stage of our research, we organized four co-design workshops pairing PwMD and robotic practitioners to explore how their early collaboration could inform accessible robot design. The procedure of the workshop was:

\subsubsection{Onboarding and Introduction:}
Participants were sent preparatory materials introducing them to the format of the workshop and sharing examples of existing public robots such as delivery robots and security robots. When the workshop started, we introduced the facilitators, and each participant, and emphasized that the two groups' insights are equally important. 

\subsubsection{Activity 1 – Conceptualize and Ideate the Robot:}
\label{method:workshop-activity1}
In the first part of the co-design session, participants were asked to reimagine sidewalk robot technology such that accessibility is centered, and broad public interest is honored. We also provided prompts including functionalities and scenarios to help them narrow down (refer to \ref{Co-design workshop activity 1}. One of the facilitators hand sketched participants' ideas as they discussed, and later asked participants for input or modifications to both ensure accuracy of representation and to promote ongoing discussion. 

\subsubsection{Activity 2 – Robot Communication \& Interaction Under Different Scenarios:}
\label{method:workshop-activity2}
\begin{figure*}[h]
     \centering
     \begin{subfigure}[h]{\textwidth}
         \centering
         \includegraphics[width=0.7\textwidth]{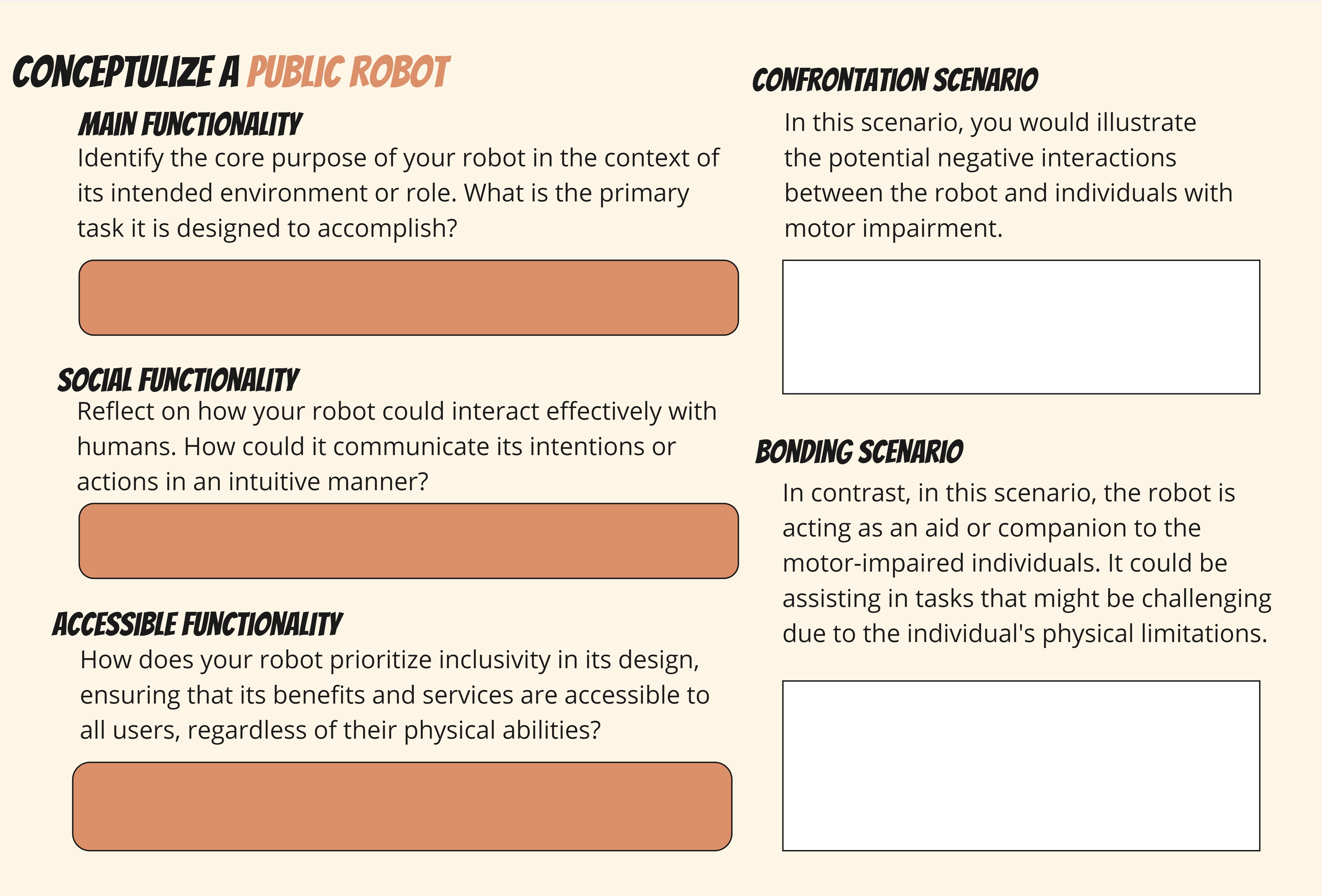}
         \hfill
         \caption{The 'Conceptualize a Public Robot' board includes a series of prompts for the first activity of our workshops. We outlined three aspects of functionality to help participants navigate to the goal. After they decided on the main idea, we encouraged them to think about negative and positive scenarios.}
         \label{Co-design workshop activity 1}
     \end{subfigure}
      % \hspace{1em} % Adjust the space to your liking
     \begin{subfigure}[h]{\textwidth}
         \centering
         \includegraphics[width=0.7\textwidth]{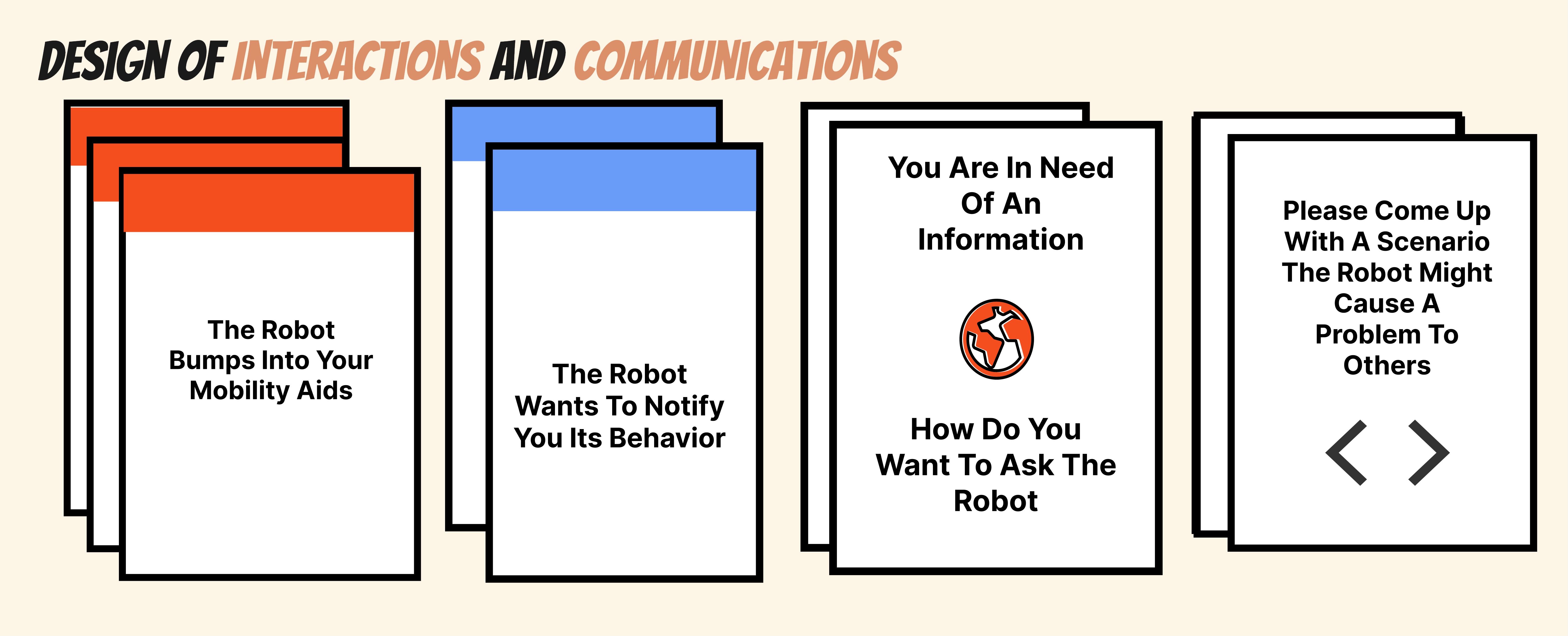}
         \caption{The 'Design of Interactions And Communications' board contains cards about both negative and positive encounters with the robots (e.g., the robot bumps into your mobility aid). Participants selected cards based on their robots. We then asked what interactions or communications they would like the robot to have.}
         \label{Co-design workshop activity 2}
     \end{subfigure}
          \hfill   
     \caption{Two elements of the design activity template that we used to facilitate the co-design workshop}
     \Description{This figure displays two images depicting the design activity templates used in a co-design workshop. The first image shows the 'Conceptualize a Public Robot' board, which contains a series of prompts guiding participants in conceptualizing a public robot, encouraging them to think about both negative and positive scenarios surrounding their main idea. The second image showcases the 'Design of Interactions And Communications' board, where participants could choose cards representing potential negative and positive encounters with robots, to further discuss desired interactions or communications the robot should have. Both boards are tools to facilitate brainstorming and discussion during the workshop.}
\end{figure*}

After participants converged around a particular robot idea, they moved to the second activity, which involved picking a card with a scenario and developing the interaction and communication of the robot (refer to \ref{Co-design workshop activity 2}). The scenarios contained situations from the first activity, as well as predetermined encounters, such as robots stuck in the road. To support the activity, we provided the interaction modalities and communication channels we used in the first interview for their reference. 

\subsubsection{Discussion and Reflection:}
Finally, participants revisited their initial robot designs, often making adjustments and reflecting on how engineers could enhance engagements with PwMD. The workshop concluded with a survey to gather feedback on its effectiveness and participants' views on creating accessible public robots.

\subsection{Data analysis}
We analyzed our data using a mixed inductive and deductive qualitative analysis~\cite{fereday2006demonstrating}. To begin,
three research team members developed a set of deductive codes by reviewing the interview and co-design study protocols. The deductive codes were mainly based on the pre-set topics in our interview questions.
For example, the interview of PwMD contains pre-set topics on interaction preferences, design factors, and attitudes toward sharing the sidewalk. Then, three team members collectively coded two interview transcripts and two workshop transcripts, combining the deductive codes and bottom-up coding strategies to build an initial codebook. The first author independently coded all remaining data, with all authors convening to discuss and resolve conflicts (e.g., missing codes, disagreement on codes). Each code, along with its corresponding interview transcript section, was sorted into an affinity diagram. This process continued until broader patterns and relationships appeared. The semantic analysis steps stopped when all team members reached a consensus around key themes, including PwMD's perceptions of sidewalk robots, practitioners' challenges for ensuring accessibility, and robot ideas in co-design.
% The final codebook is shown in Appendix \ref{apdx:codebook}.

\section{Findings: PwMD's perceptions of sidewalk robots} 
In the first section of our findings, we describe how PwMD perceive sidewalk robots, examining various design facets, interaction dynamics, and prospective deployments. 
We find three potential conflicts around the current scarcity of sidewalk space and curb cuts, robots causing PwMDs to inconveniently change their path, and an inability to communicate with robots.
We also find two themes focusing on interaction that ensures safety and usable communication modalities. 
We conclude with two themes highlighting future expectations on robots having altruistic functions benefiting disabled people and ensuring accountability and transparency of robots operations. 
% The insights were mainly drawn from the interview of PwMD but also combined comments from roboticists. 

\subsection{Concerns on Robots' Presence and Design}
\label{sec:PwMD-concerns}
\subsubsection{Potential Conflicts Over Competing Sidewalk Resources}
\label{Conflicts over Sidewalk Resources}
The \textbf{scarcity of sidewalk space} was a unanimous concern among participants, should robots fail to adjust their routes to make way for passersby. For the PwMD we spoke with, this issue significantly complicates the process of maneuvering around the robots, potentially forcing them to \textbf{resort to dangerous or inconvenient path adjustments}. P1, for instance, expressed concerns about getting trapped on the sidewalk when encountering these robots:
\begin{quote}
    ``...especially if there are also trees, and bricks that might obstruct me. Do I have any space to move around? Most likely, I would have to go down the sidewalk! But after I get down, can I come up again, or is there any way to make [the robot] move away?''
\end{quote}
A similar concern was echoed by 9 other participants, who suggested various \textbf{coping strategies}, such as backing up and leaving the sidewalk to allow the robots to pass (P1, P6, P9, P12, P14), venturing onto potentially hazardous grassy or uneven areas (P4, P9, P11), seeking shelter by residences or stores (P8), or modifying their original route (P10, P15). According to our interviewees, some of these actions could pose challenges due to one's mobility disability. For instance, once one leaves the sidewalk, it might be difficult to return. P9 mentioned how their wheelchair could easily "sink into the grass, especially if it's raining." These potential conflicts induced feelings of frustration, fear, and even anger among the participants. P3, for instance, claimed they might even \textit{“knock it over if that thing [robot] was in the way.”} 

Our interviewees also described curb cuts as another critical resource and site of potential conflict with sidewalk robots. Participants discussed that the real-world accidents, such as \cite{ackerman2019lessons}, mentioned before might not be isolated incidents. All participants acknowledged the irreplaceable value of curb cuts and voiced concerns about how robots' presence might exacerbate their existing scarcity. P14 explicitly shared their worry that \textit{``...\textbf{the curb cuts could be a point of contention} when I need the curb cut and the robot also needs the curb cut when we're trying to get on.''}  This apprehension around the loss of access to curb cuts was amplified when interviewees considered the scenario of multiple robots operating on sidewalks concurrently. 

\begin{figure*}[h]
     \centering
     \begin{subfigure}[b]{0.50\textwidth}
         \centering
         \hfill
         \includegraphics[width=\textwidth]{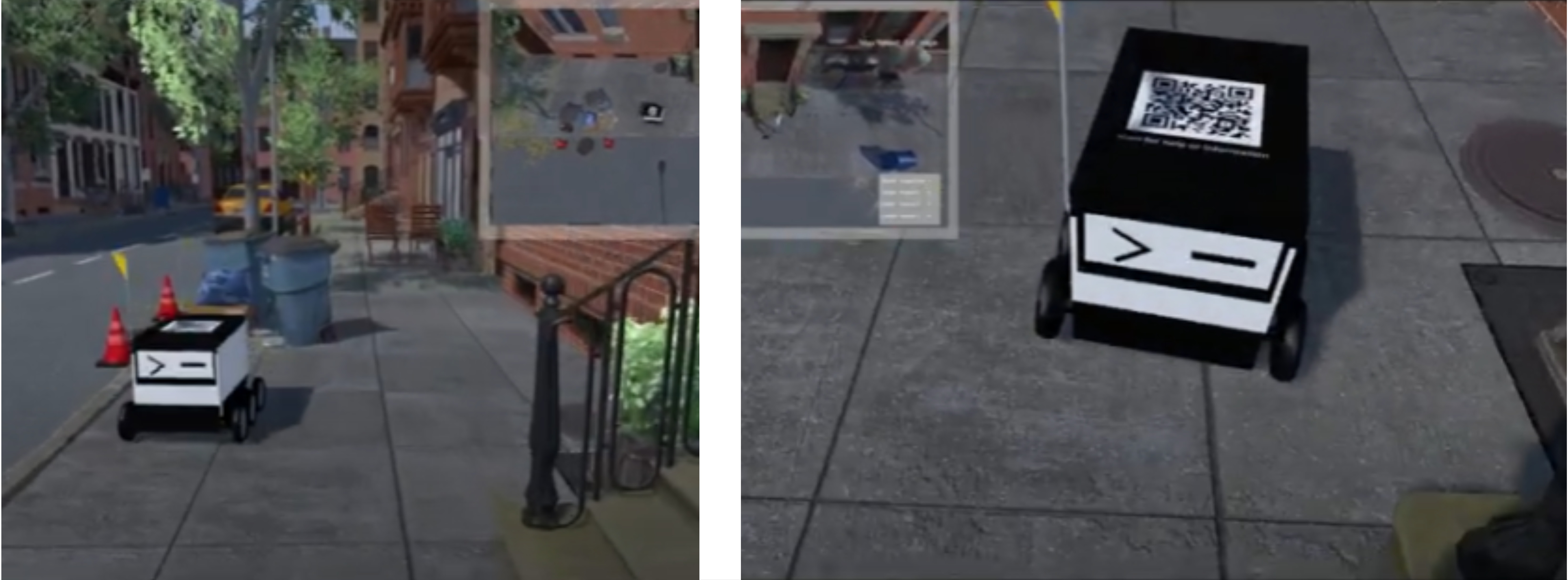}
         \centering
         \caption{A simulation video used to showcase different strategies of social navigation with sidewalk robots}
         \label{Fig:socialnavigation}
     \end{subfigure}
      \hspace{2em} % Adjust the space to your liking
     \begin{subfigure}[b]{0.35\textwidth}
         \centering
         \includegraphics[width=0.8\textwidth]{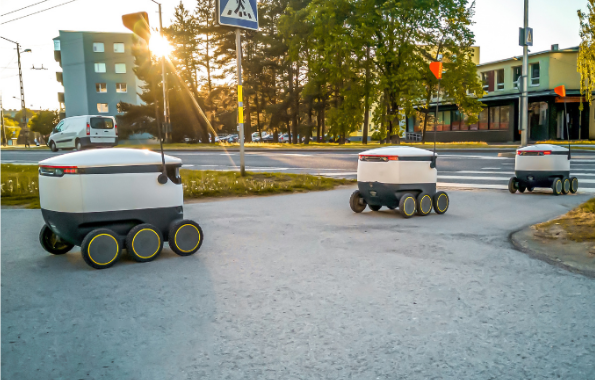}
         \caption{An image used to probe how PwMD would respond to multiple sidewalk robots}
         \label{Fig:multiple-robots}
     \end{subfigure}
      \caption{Images presented during the interviews with PwMD. The images depict possible encounters with sidewalk robot scenarios to probe design factors.}
     \hfill
     \Description{The figure contains two images presented during interviews with PwMD to explore potential encounters with sidewalk robots. The first subfigure showcases a simulation video representing various strategies of social navigation with sidewalk robots. The second subfigure offers a visual stimulus to understand how PwMD would respond to a scenario featuring multiple sidewalk robots at a crossroad. The images are meant to provoke discussions around design factors for sidewalk robots.}
\end{figure*}

When presented with an image of several robots on a sidewalk (refer to Fig.\ref{Fig:multiple-robots}\footnote{Image resource: \url{https://smudailycampus.com/1060221/news/starship-food-delivery-robots-land-at-smu/}}), an overwhelming majority of participants (14 out of 15) reacted negatively. P2 mentioned:\textit{``
...If I come across a line of small robots near the curb cut, I might choose to stop, even if I have pressing matters to attend to...''} P1, P9, and P12 shared similar \textbf{anxieties about navigating past a group of robots}.
Despite there being space between the robots in the image, P9 and P12 expressed hesitation, attributing this to the challenge of judging the adequacy of the space and the unpredictability of the robots' movements.

% Given that the sidewalk robot was a relatively novel concept for most of our participants (and the public more generally), interviewees noted that \textbf{sudden changes in their behaviors would lead to confusion}. For instance, P8 exhibited bafflement when a robot paused to give way in one of the videos depicting its behaviors, leading to speculation about potential malfunctions and concern that \textit{``the robot stops for a specific reason.''} This resulted in a reluctance to proceed, driven by fear and uncertainty. P2 expressed a similar sentiment, suggesting, \textit{``I would probably let them pass first because I am somewhat fearful.''} Such uncertainties, compounded by a reluctance to approach the robots out of causing damage, could pose significant navigation challenges for PwMD when encountering these machines in public.

\subsubsection{Sense-Making and Communicating Sidewalk Robot Behavior}
\label{sense-making}
Interviewees described how explicit communication could prevent and mitigate potential conflicts. All participants emphasized the significance of communication, and nine (P1-5, P7-9, P13) suggested that \textbf{robots should announce their presence} when turning, changing speed or trajectory, or approaching PwMD. Voice announcements were viewed as an effective and intuitive method for robots to communicate their presence and intentions on public sidewalks. P4 and P5 additionally suggested a polite, soft beep as an acknowledgment of PwMD's presence or intent to change directions. 

% P9 underscored the importance of detailed verbal communication:

% \begin{quote}
% ``[...] If someone says 'turning right', I'm probably going to instinctively move away from that corner [...]. If [the robot] doesn't stop, it should let people on the other side know that it's turning and how long it will wait.''
% \end{quote}
% P1 echoed these sentiments, suggesting that robots should guide PwMD on how to avoid a collision, particularly in potential conflict situations. 

In addition to auditory cues, visual cues such as text and icons were deemed lightweight yet effective communication channels, especially in ``crowded situations where auditory cues might be lost'' (P13,15). However, participants also highlighted some limitations of visual cues. For instance, the robot's low height might make legibility difficult, and the cognitive load associated with comprehending text might be a barrier to quick transmission of the message (P5,7). Therefore, combining text with icons could improve understanding (P7). 
% P13 also suggested using lighting to indicate changes in direction or stopping, similar to a car's signal system.

Nevertheless, P9 pointed out the potential for misinterpretation with oversimplified signals, such as confusion over who should move when a left turn icon is displayed. Auditory cues could also lead to confusion on crowded sidewalks, where the source of sounds might not be immediately identifiable.

\subsubsection{Recognition and Adaptation to PwMD} 
\label{Recognizing and adapting to PwMD}
Nine participants expressed concerns about whether sidewalk robots would be capable of effectively navigating around them, given that their mobility aids might cause them to move differently than people walking on feet. As a remedy, several participants suggested they would be reassured if \textbf{the robots show that they can recognize their disabilities} and \textbf{the robot's navigation algorithm could adapt to their mobility aids.} P8 and P12 expressed desires for robots to exhibit some form of acknowledgment that they have detected a wheelchair or sensed behavioral shifts, like moving slower than other nearby pedestrians. P4 underlined the importance of recognition as fundamental to robots learning to \textit{``treat wheelchair users respectfully''}. 
% They also suggested a user interface similar to those used in autonomous vehicles, where pedestrians' movements and relative positions are displayed, so they can be more aware of the robots' intentions and planned actions.
At the same time, our participants also raised concerns about robots' proficiency in accurately recognizing and \textbf{responding to a diverse range of mobility aids} and their distinct usage patterns, which could be challenging even for the users themselves to articulate.

% However, in our late interview with roboticists, one of the robotic practitioners (R4) strongly opposed the recognition because of ``[robots'] inaccuracies and inherent bias that exists in computer vision already''. ``If we start to tag people, it is going to get really ugly very quickly.'' thus, they argued to ``design for universal safety regardless of people's disabilities''

\subsubsection{Robot Design Factor Tensions}
\label{HRI design factor tensions}

When presenting the design factors, there was always no unified preference --- each option carries its rationale and limitations. For example, in terms of form, large robots were more likely to be viewed as \textbf{obstructing sidewalks and invoking fear} (P2). Smaller robots elicited less fear and were thought of as taking up less sidewalk space, but they presented a different issue of potentially trapping pedestrians. P8 observed that \textit{``(Wheelchair users) may not see the smaller robot if not spotted from a distance.}'' P1 expressed concerns about the difficulty of interacting with shorter robots: 
\begin{quote}
I feel that the (small robot) is less user-friendly because I can at least touch the top of the taller one.[...] taller one might have operational or interactive functions on top, so I can effectively communicate or signal that it has impacted me[...]
\end{quote}

However, in contemplating whether the robots could halt their movement to allow someone to pass, P1, P3, and P13 highlighted a key concern that the \textbf{robots' stopping points might coincide with the trajectory of their wheelchairs.} To avoid the stationary robot, they would have to adjust their own paths. This adjustment is not a straightforward task; for PwMD, especially those relying on manual wheelchairs, altering direction involves physical effort and precise maneuvering, which can be time-consuming. 

% Robot form also affected \textbf{signal visibility and interaction accessibility}. Six participants expressed concern that information displayed on the robot, such as text or images, might be challenging to see. 
% \begin{figure}[h]
%     \centering
%     \includegraphics[width=\textwidth]{Pictures/HRI_design tensions.png}
%     \caption{This figure depicted three design tensions from our interview findings. In terms of size, small robots were perceived to be more perceivable. While might be intimidating, large robots are perceived to be more reachable  }
%     \label{Design tensions}
% \end{figure}

\subsection{Designing Interaction with Sidewalk Robots}
\subsubsection{Purpose of the Interaction with Sidewalk Robots}
\label{Purpose of the interaction with Sidewalk Robots}
Considering the potential conflicts and confusions presented above, all participants underscored the importance of \textbf{interactions with sidewalk robots to ensure safety.} Critical situations like competing for curb-cut access or \textbf{evading dangerous areas necessitate effective interaction with the robots}. Voice interaction was considered the most effective mode in these cases, with P6 stating they would vocally command the robot to move. P8 highlighted their comfort in expressing their needs and emotions to the robot: 
\begin{quote}
If the robot makes some movement that could lead me to a problem[...] I will say that can you choose another way? Or is there another option because I'm not comfortable?  
\end{quote}

Access to sidewalks was a significant concern for participants, emphasizing the need to \textbf{command the robot to leave the sidewalk} through a button or voice command. P9 elaborated, \textit{``The biggest thing for any sidewalk robot is, there [must be] a way for it to get off the sidewalk...''} This sentiment was echoed by P1, P3, and P10, as they felt it could alleviate potential bottlenecks of the curb cut and better accommodate narrower sidewalks.

While participants generally felt less need to interact with robots in non-conflict scenarios, they acknowledged that such \textbf{interactions could alleviate anxiety}. Some participants even expressed a desire for casual, human-like interactions with the robots. For instance, P4 said that when a robot passes close by, they would like it to say \textit{``excuse me''} much like another pedestrian would.

\textbf{Addressing robot malfunctions} is another circumstance where participants underlined the need for effective interactions. P7 and P8 called for a feature to directly contact operators to address arising issues. However, they also acknowledged that operators might not always be immediately available. In such scenarios, the robots should be programmed to \textbf{deliver clear, comprehensible instructions to the public}. This way, bystanders might to more able to assist in mitigating the situation. The guidance could encompass steps to temporarily disable the robot and safely move it out of the way, effectively preventing the malfunctioning robot from becoming an obstacle on the sidewalk.

\subsubsection{Exploring the Usability of Interaction Modalities in Public Settings}
\label{Usability of different interaction modalities}
In pursuit of efficient interactions, we examined the \textbf{usability of six potential interaction modalities}. Voice interaction and touchpads emerged as popular choices. P5, P6, and P9 preferred voice interaction for its convenience and non-disruptive nature, as it doesn't necessitate pausing current activities. However, P4 raised concerns about the \textbf{reliability of voice recognition}, especially regarding accents, fearing this could lead to miscommunication. They preferred a screen-based interaction for its clarity and definiteness, as \textit{``the UX[user experience] design can give clear instruction and meaning without ambiguity.''} The merits of physical interactions were acknowledged by P1 and P3, who argued that buttons could \textbf{distill complex commands into simple, easily understood interactions.} 

However, participants also cautioned that direct \textbf{contact with the robots might interfere with their personal affairs}, with P10, emphasizing, \textit{``If interaction forces people to stop and spend a significant amount of time, the robots will quickly fall out of favor.''} Balancing the need for effective communication and personal convenience, P8 and P10 favored app-based solutions. They appreciated the non-disruptive and streamlined nature of this modality.P1 also spoke about how they use gestures to communicate with other vehicles and reasoned that if robots could accurately recognize these gestures, this modality could prove helpful in adjusting social distance or directing robot movements.

% \begin{figure}
%     \centering
%     \includegraphics[width = \textwidth]{Pictures/Interaction modalities.png}
%     \caption{Interaction modalities- I will adjust the font and content later}
%     \label{fig:enter-label}
% \end{figure}

\subsection{Expectations and Aspirations for Accessible Public Robots}
\label{sec:expectation}
\subsubsection{Functionalities and Information Beneficial to the Disabled Community}
\label{sec:altruistic functions}
80\% of the participants expressed interest in using delivery robots but felt that the \textbf{delivery functionality alone was insufficient for serving and benefiting the disabled community.} Alternate forms of sidewalk robots, such as snow-clearing and safety robots, were perceived positively. For instance, while P3 was not in favor of delivery robots, they found the concept of snow-clearing robots intriguing, especially due to the difficulties posed by snowy conditions. Other participants cited the challenges icy roads present during winter. P11 remarked, “\textit{If the robot could alert me about hazardous roads, it might embolden me to venture out during winter}.” P2 and P6 floated the idea of multi-functional robots that primarily handle deliveries but could also provide other functionalities, enhancing their acceptability. For example, P2 argued that \textit{``Since it's a delivery robot, it might also bring some first-aid medicine that could save people's lives''} and thought it could contribute to positive public perception. 
% This underscores the importance of understanding the diverse needs of people with disabilities in various public settings when designing robotic functionalities.

Discussing the potential data types beneficial for detecting road conditions, the concept of \textbf{robots disseminating road information was generally welcomed}, aligning with current trends in mobility technologies (refer to Section \ref{r1-inaccessible}).  P4, for instance, commented that he had already installed a local app that reports roadblocks and believed that the robot ``\textit{can definitely help more because they are patrolling all the time and the data will be more up-to-date}.'' Pothole detection was suggested by P1, P3, P6, P7, and P8, with P8 adding that the depth of potholes is another crucial parameter for PwMD since they often do not notice them until ``\textit{they actually get stuck by the potholes}.'' The identification of roadblocks and obstacles like sticks and pipes was also recognized as contributing to the navigation experience of PwMD. 

\subsubsection{Ensuring Robot Accountability and Transparency}
\label{sec:ensuring robot accountability}
Beyond expanded functionalities,  all the participants emphasized the crucial role of \textbf{handling technical failures to ensure safety.} P3, in particular, stressed that while technical failures are inevitable, ``\textit{there should be ways to mitigate}'' issues. As described in the interaction section (refer to Section \ref{Usability of different interaction modalities}), alternative modalities were suggested by participants to counterbalance the effects of one function failing. This call for a failure-handling function to maintain robot communication was also echoed by P1, P3, and P6. P1 proposed the inclusion of emergency buttons, so the robot could move itself to a safer location when it becomes obstructed. 

When operating normally, participants expressed the expectation for the robot's interaction to remain stable and accountable, notwithstanding environmental influences. If certain interaction modalities fail, the robot should explicitly communicate this. P4, for example, stated that if the voice system is less accurate in recognizing their sounds, the robot should be transparent about this and provide instructions on how to use other interactive methods, such as a touchpad, to fulfill their needs. The desire to understand how to interact with the robots and their performance parameters under different conditions was voiced by P1, P4, P7, and P10.

% The frequency with which individuals might encounter robots was a topic often discussed by participants (10 out of 15). Drawing a parallel to Amazon trucks, P13 conjectured, ``\textit{the robots would go all over the place}.'' Privacy concerns arose in relation to the robots' cameras and sensors.  P3 voiced a concern, ``\textit{If the robots were everywhere, I would worry about the device being hacked.}'' 

Twelve participants also advocated for \textbf{engaging in conversations with governments and robot companies.} P14, a wheelchair user and software engineer, believed that robot companies should make changes to their processes to incorporate accessibility, including regularly talking to PwMD. P10 remarked that governments should be more involved in the oversight and deployment of robots in public: ``\textit{I think the technology can only be beneficial if the governments introduce them in the right way}.''

\section{Findings: Navigating the Terrain of Organizational Practices – Challenges and Potential}
Our findings so far suggest current robotic practices may not sufficiently account for accessibility. Thus, in this section, we examine the views of robotic practitioners from both industry and academia who reflect on organizational challenges hindering accessibility, and how they might better ensure inclusive practices by integrating conversation with people with disabilities and disability-oriented design thinking early into the robot design process. 
In the context of PwMD's needs, we also investigate strategies for deploying and operating public service robots by collecting feedback from diverse groups of stakeholders in real-world settings and in developing plans to fix emergent issues. Furthermore, our findings indicate roboticists suggestions for comprehensive regulations on robot speed, weight, and interaction design that may lead to more accessible robots by design.

% \begin{figure}
%     \centering
%     \includegraphics[width=\textwidth]{Pictures/Findings from Practitioner.jpg}
%     \caption{ Twelve findings, grouped into four categories, from the interviews with robotic practitioners}
%     \label{fig:enter-label}
% \end{figure}

\subsection{Organizational Practices for Infusing Accessibility into Robotic Development}
\label{organizational challenge}
\subsubsection{Obstacles to Guarantee Accessibility}
Industry practitioners (R4-7) shared their experiences on how robotic companies often \textbf{overlook accessibility}. R5, for example, drew attention to feature prioritization metrics their company followed, noting, \textit{``[the company] always focus[es] on high impact, low effort tasks first''}. Unfortunately, when the "impact" is measured by the number of affected individuals, accessibility initiatives can be sidelined, R5 explained. Even when robotics companies anticipate interactions with disabled people, representatives may point to limited funding or time as a reason for not considering disabled people's needs. R4 expressed this candidly, and R5, R6, and R7 echoed the sentiment:

\begin{quote}
The problem is that most of the companies developing robots are startups, and they don't care, or they don't have the time to be able to do the proper compliance steps[...] It's just hard and expensive. And so a lot of startups, either don't know it exists or they do know it exists[...] It's not worth doing because they would run out of money before they could make their first robot (R4). 
\end{quote}

As a result, instead of proactive measures, robotics companies might `\textit{`prioritize it only after a problem occurs (R5)''}  and \textit{``hope they can convince the regulators to not shut down their companies [after the problem] (R6).''} 
However, most of our participants shared the arguments that \textbf{accessibility issues are harder to tackle if not integrated from the start}. R5 and R7 shared this point based on their previous experience facing compatibility problems. R7 reported problems they faced when retrofitting to improve the user interfaces as ``complex hardware, rudimentary software, interoperability problems.'' R6 referred to their experience working on safety and asserted that if you fail to consider safety metrics early, and \textit{``only do it as a checkbox later, you're almost certainly not going to pass.''} 

% Our interview with robotic practitioners indicated that robot companies might neglect accessibility because of the limited resources, and this would only make the problems hard to resolve later. 

\subsubsection{Importance of Considering People with Disabilities}
All robotics practitioner participants believed that \textit{co-design and early inclusion of PwMD and other PWD is necessary.} R5 emphasized \textbf{the importance of directly engaging people with mobility disability}. They referred their experience pitching to public sectors, and noted that showing the response from PwMD could support their illustration and get approval. R8, an HRI researcher, championed direct involvement over empathetic design for its potential to ``minimize discrepancies and understand genuine challenges.''
R5's experience revealed how initial product ideas often undergo numerous modifications. And ``engineers tasked with these changes might lack an accessibility perspective, potentially bringing more inaccessible features.'' R4 believed that to have the right voice, there is a need to ``have disabled workers in the room'' which will ``automatically build solutions that have some weight [around accessibility] to them.'' 
  
Additionally, our participants also noted an \textbf{incentive for robotic practitioners to embrace accessibility} since avoiding accessibility can harm reputation---a point raised by R2, R4, R6, and R7. R7 stated:

\begin{quote}
If you're asking, well, can we afford to do accessibility? That's really the wrong question. The question is more like, are we an ethical organization, and do we want to make decisions with an ethical mindset as a rule? (R7)
\end{quote}

R1, R4, R5, and R7 endorsed the notion that \textbf{considering accessibility improves product design and benefits everyone} and commented that designing for people with disabilities can improve acceptance of robots among broader groups of people. ``If the robot companies evaluate the impact of accessible features,'' said R5, ``they would see how broadly it can be useful.'' R4 recounted how designing for the elderly indirectly improved robot communication, turning it into a competitive advantage. 

Ultimately, our participants believed that prioritizing accessibility offers a plethora of benefits that robotic companies often undervalue, such as preventing later costly modifications and improving their products. However, to achieve these benefits, there is a need for \textbf{broad-ranging collaborations both within and outside these organizations}. To comprehensively examine accessibility, R6 envisioned a multifaceted team, integrating UX researchers, HRI specialists, and professionals well-acquainted with the robot's operational milieu. Addressing legal hurdles, for example, would require team members with knowledge of local and federal regulations. R2 stressed that collaboration should span the spectrum of robotic development because it ``bridges the divide between user requirements and technological solutions.'' Further, deployment in public spaces brings to light the crucial need for collaboration with policymakers. R5 pointed to, what they described as, impediments introduced by government regulations, which blocked the roll-out of particular features. Across our interviews, participants advocated for a comprehensive approach, combining legal insights, research expertise, and safety awareness, to achieve an inclusive, safe, and user-friendly robotic landscape.

\subsection{Deployment and Operation of Public Robots}
\label{deploy public robots}
Designing robots for public venues is complicated, and practitioners in our interviews discussed the multiple challenges inherent in this domain. R2 emphasized the \textbf{diverse reactions of bystanders}, pondering over the dilemma, \textit{"How can a robot engage without disrupting regular activities in public spaces?"} Echoing concerns expressed by PwMD (refer to Section\ref{sense-making}), R2 noted that a robot's unfamiliar presence might cause individuals to halt their activities, suggesting a need for \textbf{clearer robot communication}. This would require modeling human perception to the robot behavior (R1), and versatile digital, and physical engagements (R8). R1, an HRI researcher, articulated a longstanding challenge:
\begin{quote}
Appropriately modeling human behavior[to the robots]...is a huge challenge that we still need to do a lot of work [...] to have robots actually behave the way we want them to and expect them to.
\end{quote}
Thus, prior to any large-scale robot deployment, practitioners argued that it is pivotal to \textbf{discern how PwMD interacts with robots in real-world settings}. R2 and R6 pointed out potential disparities between what users say and how they act. Achieving this understanding would require the development and use of research methodologies that uncover how people interact with robots, such as 
% delving into the root causes of user discomfort by contextual testing(R5-7), 
ethnographic observation (R4), and establishing robotic simulation platforms to test and tweak robot behavior (R2, R8).

Designing robots for complex public spaces also requires an understanding of the environment and its possible impacts on human-robot interactions. As highlighted by R6, even seemingly defined environments, such as a museum with flat floors, might surprise roboticists. Such a place can have ``variations in flooring—shiny, matte, black, white, all things that make robots upset... and exponentially increase the complexity of running safety analysis." In even more complex and unpredictable public spaces, R4 remarked that robots will inevitably fail, which makes designing for failure cases and immediate damage control crucial to mitigating negative or dangerous interactions. 

Offering a provisional solution to such failures, R2 suggested robots be equipped to \textbf{guide the public in fixing emergent malfunctions, and apologize for mishaps}, an idea corroborated by prior research \cite{desai2013impact} and echoed by R3. However, as R6 emphasized, failure control might not suffice in extreme situations, such as a robot running into a toddler. Further, R6 asked, "How effective would human intervention be in these scenarios?" Drawing from their experience in safety assessments, R6 elaborated, "\textit{Merely making marginal safety enhancements to a system capable of inflicting potential harm doesn't fundamentally alter the risk equation.}" R6, therefore, advocated for stringent regulations overseeing public robots, up to and including potential product recalls when necessary.

\subsection{Regulatory Frameworks and Guidelines for Robot Accessibility}
\label{need of regulation}
Given the intricacies of the robotics industry and its interactions within public spaces, five participants(R2, R4-5, R6, R8) also shared \textbf{the need for more regulations and guidelines} for all robotic development stages. R6 observed that while certain robotic firms have established safety protocols, emerging sectors within robotics lack the institutional "muscle memory" or corporate culture for structuring design processes with safety and accessibility at the forefront. The absence of regulations and guidelines has implications; for example, roboticists may be inclined to prioritize novel features over core concerns of safety and accessibility. This leads to a potential introduction of features that may compromise safety or inclusivity. R2, R5, and R7 also called out the need for well-defined design standards, particularly given a lack of concrete, user-friendly accessibility guidelines for robotic design. Drawing a parallel to web accessibility standards such as Web Content Accessibility Guidelines (WCAG)\footnote{\url{https://www.w3.org/TR/WCAG20/}}, R7 underscored how integrating accessibility checking into workflows to evaluate and address accessibility issues can empower practitioners.

Several participants requested action for overarching regulations, possibly spearheaded by governmental bodies. Current regulations mainly focus on parameters like weight limits and speed \cite{villaronga2019robots}, but R8 argued that there's a broader spectrum to examine. Interviewees proposed rules like a robot being required to maintain a certain distance from people (R2), automatically replanning when it detects people approaching curb cuts (R3), following the social norm of staying on one side of the road in most cases (R3), and restrictions on the concurrent number of robots on a single road (R3). 
Other ideas included changes to the built environment that would impact how robots and people interact. For example, R2 and R8 shared ideas of dedicated pathways for robots on sidewalks, especially if their prevalence surges. R8 visualized a synergy with smart infrastructure:
\begin{quote}
When the robot is coming, we will know that it's coming, and it's going to use this space. But at the other time, we can definitely use the whole space when there is no robot... I do not even have to look at the robot. I know, because of the lighting that is displayed on the road or maybe projected on the road, the LEDs (R8).
\end{quote}
However, our discussions primarily revolved around regulating robotic behavior rather than changing the environment, likely due to participants' HRI backgrounds. 

% Also, considering their call for directly engaging PwMD, it is imperative to explore
% how does an early conversation pairing them bring new insights to the accessible public robot design?

% A broader Discussion section (refer to Section\ref{discussion:guildlines}) delves into potential future regulations intertwined with disability laws, urban policies, and autonomous device legislation.

\section{Co-designing public service robots, pairing PwMD and robotic practitioners}
Across both sets of interviews, roboticists and PwMD called for a need to collaborate with each other. 
PwMD argued that public sidewalk robots should bring more value and designers should work to counter the potential negative effects on their navigation (Section \ref{sec:expectation}). Furthermore, PwMD desired to be more informed throughout the robot design process (Section \ref{sec:ensuring robot accountability}), while roboticists believed that early conversations with PwMD could help prevent accessibility problems later. To explore what kinds of ideas PwMD and roboticists might conceptualize together, we held four co-design workshops including our PwMD and roboticist interviewees. 

In the workshops, we saw PwMD imagine robots being deployed in ways to alleviate their accessibility challenges as well as serve the public good. PwMD proposed novel functions of robots, and roboticists elaborated on the concepts by assessing their technical feasibility. Teams considered the physical factors (e.g., size, colors, morphology), communication systems (e.g., screens, voice commands), and interaction dynamics (e.g., movement patterns, approach behaviors) of potential robots. 
%Overall, PwMD and roboticists collaborated respectfully, creatively, and practically to conceptualize a future of accessible public robots. 

\subsection{Overview of the Generated Robot Ideas}
\label{co-design:outcome and ideas}
Below, we present each of the four robot ideas from the collaborations.
To contextualize each idea, we describe the team composition and how each team interacted with one another\footnote{We used pseudonyms for all participants. Detailed information of each participant can be found in Table \ref{Workshop-table}}.
These ideas focus on robots that could support PwMD and include a cargo-carrying robot, a robot to grab and hold groceries in a store, a crosswalk guide robot, and a snow plow robot for clearing sidewalks.

\begin{figure*}

\noindent
\textbf{Workshop 1 (W1)}: Cargo Carrier \\
\rule{\textwidth}{0.1pt} % Break line

\noindent
\begin{minipage}[t]{0.17\linewidth}
\textit{Lily:} Cane user\\ 

\textit{Bella:} HRI researcher
\end{minipage}
\hfill
\begin{minipage}[t]{0.75\linewidth}
Lily, a cane user, and Bella, an HRI researcher, collaborated to design an assistive cargo-following robot (refer to Fig.\ref{co-design:Cargo Carrier}). To serve broader populations, they envisioned the robots being available for rent to any PwMD who might need them. They considered ideas like height adjustability and showing an avatar on the robot screen to display whose cargo it is carrying. 
% (refer to \ref{I will add one image}) 
Bella introduced voice and smartphone control opportunities while noting concerns such as the robot alarming other public users. Both acknowledged the robot's accessibility challenges and proposed features, including self-parking and secondary user recognition. 
\end{minipage}

\vspace{5pt}

\noindent
\textbf{Workshop 2 (W2)}: Grocery fetcher \\
\rule{\textwidth}{0.1pt} % Break line
\noindent
\begin{minipage}[t]{0.17\textwidth}
\textit{Adrianne:} \\Power chair user \\

\textit{Michael:} CEO of a \\robotics start-up
\end{minipage}
\hfill
\begin{minipage}[t]{0.75\textwidth}
Michael and Adrianne imagined a robotic shopping aid (refer to Fig.\ref{co-design:Grocery Fetcher}). Michael is an HRI researcher and the current CEO of a robotics startup. Adrianne uses a wheelchair, making it hard to reach high shelves to collect items at the grocery store. Moreover, she also has limited hand dexterity. Together, Michael and Adrianne conceptualized a robot to help her fetch items at the store or other places when she needed to do so. Michael led their conversation by asking questions about Adrianne experience as a disabled person and the obstacles she faces, and together, they developed the idea for a robot that sought out things using a perception system, grabbed items with arms, and could be integrated onto a shopping cart. 
\end{minipage}

\vspace{5pt}

% Workshop 3
\noindent
\textbf{Workshop 3 (W3)}: Crosswalk guide \\
\rule{\textwidth}{0.1pt} % Break line

\noindent
\begin{minipage}[t]{0.17\linewidth}
\textit{Zac:} Power chair user \\

\textit{Nora:} HRI researcher \\
\end{minipage}
\hfill
\begin{minipage}[t]{0.75\textwidth}
Zac and Nora co-developed a guide robot for safe street crossings (refer to Fig.\ref{co-design:Crosswalk Guider}). Originating from Zac's challenges navigating in traffic, the design catered to a broad demographic, including dog walkers and elderly folks. Responding to Zac's concerns about accidents and pedestrians, Nora focused on the robot's core safety functions: vehicle detection, pedestrian status updates, and driver alerts. They added emergency responses and car impact resistance. Inspired by Zac's vision of connected devices, they believed the robots could have the capability to communicate with autonomous cars and halt them before they could hit someone. 
\end{minipage}
\vspace{5pt}

\noindent
\textbf{Workshop 4 (W4)}: Snow plow \\
\rule{\textwidth}{0.1pt} % Break line

\noindent
\begin{minipage}[t]{0.17\textwidth}
\textit{Will:} Power chair user \\ 

\textit{Emma:} \\Robotic engineer
\end{minipage}
\hfill
\begin{minipage}[t]{0.75\textwidth}
Will and Emma collaborated to conceptualize a sidewalk snow plow robot~(refer to Fig.\ref{co-design:Snow Plow}). Emma is an HRI researcher and engineer. Will is a lawyer and a power wheelchair user with limited hand and arm dexterity. They designed a robot meant to plow snow in extreme weather and late at night when it could be dangerous for humans to work. They imagined this technology as a method to keep streets clean and improve the efficiency of plowing snow while having minimal impact on human workers. 
\end{minipage}
\rule{\textwidth}{0.1pt} % Break line

\vspace{2pt}

\end{figure*}

\begin{figure*}[h]
     % \centering
     \begin{subfigure}{0.4\textwidth}
         % \centering
         \includegraphics[width=6cm, height = 4cm]{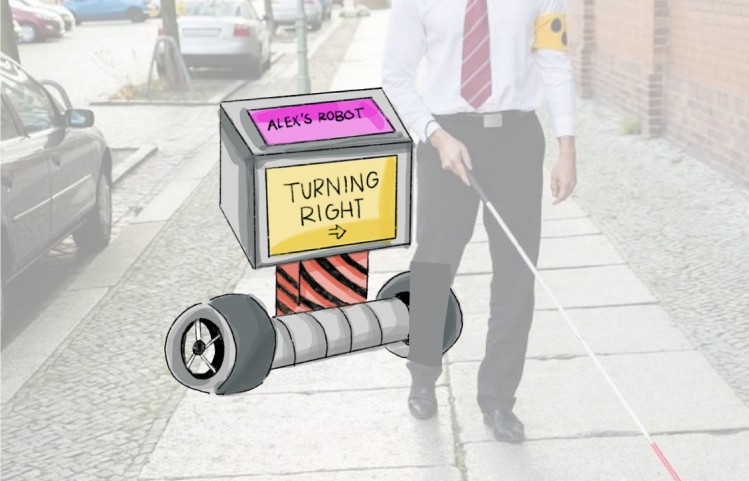}
         \caption{Cargo Carrier Robot designed in Workshop 1}
         \label{co-design:Cargo Carrier}
     \end{subfigure}
     % \hfill
     \hspace{2em} % Adjust the space to your liking
     \begin{subfigure}[b]{0.4\textwidth}
         % \centering
         \includegraphics[width=6cm, height = 4cm]{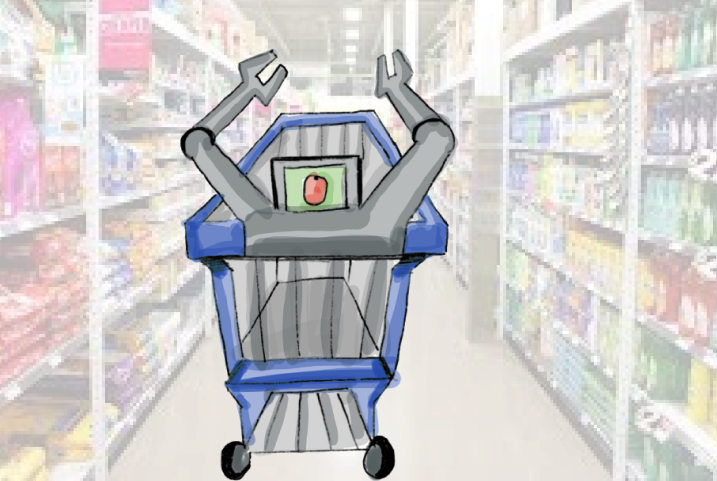}
         \caption{Grocery Fetcher Robot designed in Workshop 2}
         \label{co-design:Grocery Fetcher}
     \end{subfigure}
     \hfill
     \begin{subfigure}[b]{0.4\textwidth}
         % \centering
         \includegraphics[width=6cm, height = 4cm]{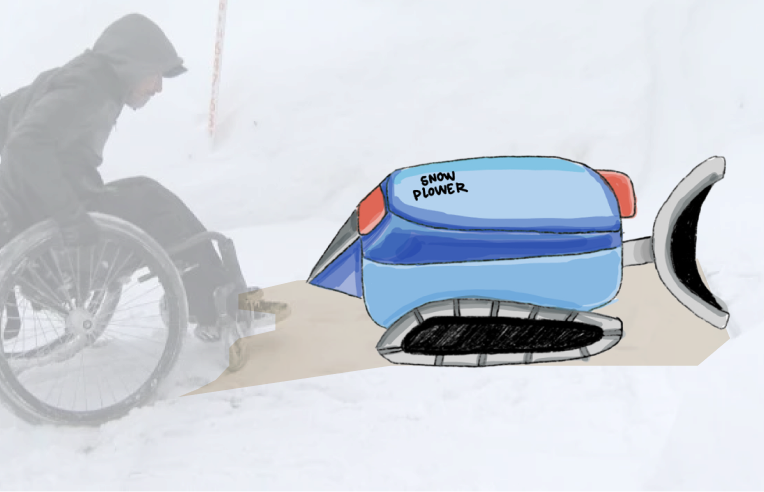}
         \caption{Crosswalk Guider Robot designed in Workshop 3}
         \label{co-design:Crosswalk Guider}
     \end{subfigure}
     % \hfill
     \hspace{1em} % Adjust the space to your liking
     \begin{subfigure}[b]{0.4\textwidth}
         \centering
         \includegraphics[width=6cm, height = 4cm]{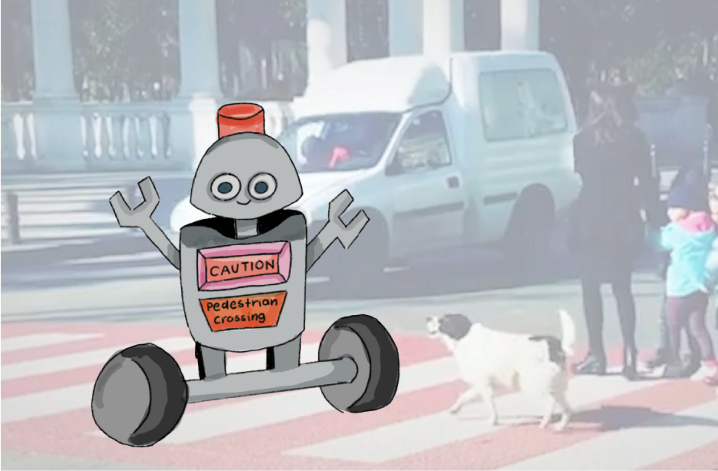}
         \caption{Snow Plow Robot designed in Workshop 4}
         \label{co-design:Snow Plow}
     \end{subfigure} 
        \caption{Final illustrations of the robot ideas from the four co-design workshops. The illustrations have also been photo-edited to depict the public venue they operate within in the background.}
        \label{W4-robot}
        \Description{This figure showcases the final illustrations of robot ideas conceived in four different co-design workshops. The visual consists of four subfigures, each displaying a robot concept with a specific function. In the first subfigure, labeled 'Cargo Carrier Robot designed in Workshop 1', a robot designed to carry cargo is displayed, situated in an appropriate public venue background. The second subfigure presents the 'Grocery Fetcher Robot designed in Workshop 2' in a similar setup. The third displays the 'Crosswalk Guider Robot designed in Workshop 3', and the fourth shows the 'Snow Plow Robot designed in Workshop 4', each situated in a background depicting the public venue where they would operate. The illustrations provide a visual representation of the robots alongside their intended functional environments.}
\end{figure*}

% \subsection{Collaborative Process and Decision-making}
% \label{Co-design: Collaborative process}
% \subsubsection{Conflicts and Knowledge Gaps}
% \label{Co-design:conflicts}
% In the workshops, we sought to elevate the needs and opinions of PwMD, facilitating the conversation that they were asked to comment first in response to prompts of designing a service robot for use in public (refer to Section\label{methods}), and encouraged them to share their expertise throughout.  Although each group had different processes for conceptualizing their robot designs, each group eventually converged on a robot that would be assistive to both PwMD and beneficial to the broader public.

% The central conflict observed during the co-design sessions was balancing the \textbf{accessibility needs brought forth by PwMD and the technical perspectives brought forth by the roboticists}. This tension bore out differently across each workshop—sometimes pertaining to a conflict between accessible features and their technical feasibility or an interaction preferred by PwMD that roboticists doubted being possible. 

\subsection{Uncovered Accessibility Needs and Concerns}
\label{co-design:uncoveraccessibility}
As part of our protocol, we asked workshop teams to center the needs of PwMD in creating their robot concept.
This led to designs that focused on specific needs for PwMD such as carrying or reaching items independently (W1, W2) and providing increased safety on public walkways (W3, W4). 
Throughout the co-design engagements, there were instances where teams needed to work through conflicts around robot features desired by PwMD and the technical feasibility of implementing such features (frequently emphasized by roboticists).
These disagreements would often lead to more accessibility needs and concerns being uncovered.
To illustrate such conflicts and how they realized more detailed accessibility issues, we present two vignettes from the workshops. 
% we reflect on opportunities embedded in the paired co-design and how this process challenges current HRI practices and offers valuable alternatives.

\noindent\textit{Vignettes 1: Why Opt For a Robot if it Can't Outperform a Human Assistant?}

\begin{quote}
    In Workshop 2 (W2), Adrianne, a power chair user with limited hand dexterity, expressed a desire for assistance in fetching groceries. When Michael, a roboticist, sought to understand how long Adrianne's grocery fetching task would take, he learned that with human assistance, Adrianne could fetch five items in approximately 20 minutes. "But would a robot match a human's efficiency?" Adrianne asked. "Not a chance," Michael admitted frankly, acknowledging the technical reality that robots would likely lag behind humans in terms of efficiently navigating a grocery store and locating items. "My life is about taking things slowly. It's fine if it takes three times longer," Adrianne responded. She shared feelings of being hesitant to inconvenience others when she needed assistance in the store. Adrianne explained that a robot might circumvent such social discomforts and support her autonomy. Through this exchange, the core value of the robot Adrianne proposed became clear to Michael: not to offer mere efficiency, but the prospect of furthering her autonomy. 
\end{quote}

% Throughout our workshops, there were moments when the roboticists questioned certain proposed designs because of feasibility constraints. 
During the co-design sessions, the PwMD and roboticists often came to a compromise on technical features, but there were also cases where PwMD remained unwavering in what they believed were their most important requirements.

\noindent\textit{Vignette 2: I Value Your Concerns, But This is What I Need As a Cane User}
\begin{quote}
     In W1, Lily, a cane user, and Bella, an HRI researcher, decided to conceptualize a cargo-carrying robot. Lily immediately proposed that the device be built to ``adjust to different heights for people to easily get access to it'' because of her inability to bend over. She elaborated:
    \textit{``If I need to press a button or tell it to do something, it's important to me that I don't have to really stretch up or really bend down to do that because that's something that would catch me off balance somewhat.''}
    Bella scrutinized this functionality, worrying that ``it might be terrifying to other groups like elders'' when changing shape. Although Lily admitted there was a possibility that a robot adjusting its height could be disturbing, she insisted it would ``be useful for broader groups of people with mobility disability like wheelchair users''  
    This feature later became one of the main attributes of the robot because Lily suggested various heights the robot should cater to, taking into account situations where she might be using a wheelchair or standing with the support of canes. She felt that she was shorter than the average person. 
    
\end{quote}
Across the Workshops, teams discussed a variety of accessibility needs that their robot concepts aimed to address.
While the true feasibility of each concept would require extensive testing, we believe that each brings up more needs that could be generalized to other robots for public use.
The following summarizes the core accessibility needs of each workshop, grouped by functionalities and interactions.

\begin{figure*}

\textbf{Table 4: Overview of User Needs and Interactions for Four Different Robotics Workshops}

\noindent \textbf{Workshop 1: Cargo Carrier Robot Needs}

\begin{table}[H]
    \centering
    \begin{tabular}{p{0.48\textwidth}p{0.48\textwidth}}
\hline
\centering\textbf{Functionalities} & \centering\textbf{Interactions} \tabularnewline
\hline
\begin{itemize}[leftmargin=*]
    \item \textbf{Carry heavy cargo} for users who may lack the ability to do so themselves.
    \item \textbf{Self-park to avoid obstructing paths} when the user enters an area where it cannot maneuver, such as near a steep ramp.
    \item \textbf{Adjustable height to accommodate people using mobility aids}, allowing use by those in non-normative height ranges.
    % \item \textbf{Signpost robot behavior} such as turning and stopping to allow PwMD to have time to respond, considering their potential movement latency
\end{itemize}
&
\begin{itemize}[leftmargin=*]
    \item \textbf{Limited control access for other sidewalk users} to move it out of the way when it is unaccompanied.
    \item Users should \textbf{be able to command the robot via voice} to prevent physical strain.
    \item \textbf{Smartwatch control option for users, like Lily, cannot use a phone while moving}, to facilitate use when standing, especially when voice recognition is unreliable.
\end{itemize}
\\
\hline
\end{tabular}
\end{table}

\noindent\textbf{Workshop 2: Grocery Fetcher Robot Needs}

\begin{table}[H]
    \centering
    \begin{tabular}{p{0.48\textwidth}p{0.48\textwidth}}
\hline
\centering\textbf{Functionalities} & \centering\textbf{Interactions} \tabularnewline
\hline
\begin{itemize}[leftmargin=*]
    \item \textbf{Pick heavy and high-placed items using a robotic arm} to assist people with limited hand dexterity and wheelchair users.
    \item \textbf{Integrate the body of robots into shopping carts} instead of having independent ground to not take extra space (Refer to Fig.~\ref{co-design:decision-1}
    % \footnote{Image resource: https://spectrum.ieee.org/fetch-robotics-introduces-fetch-and-freight-your-warehouse-is-now-automated}
    ).
    \item \textbf{Accompany users consistently} so that they don't need to request human assistance when none is available.
    % \item 
\end{itemize}
&
\begin{itemize}[leftmargin=*]
    \item \textbf{Understand natural language} when users are seeking items for flexibility \textbf{but also provide a touchpad} for communication when users might be hesitant to speak out loud.
    \item Robot should \textbf{be courteous and explicit about each action it takes}, so users can follow.
    \item \textbf{Indicate the items on the screen} to confirm with users that it understands the needs.
\end{itemize}
\\
\hline
\end{tabular}
\end{table}

\noindent \textbf{Workshop 3: Crosswalk Guide Robot Needs}

\begin{table}[H]
    \centering
    \begin{tabular}{p{0.48\textwidth}p{0.48\textwidth}}
\hline
\centering\textbf{Functionalities} & \centering\textbf{Interactions} \tabularnewline
\hline
\begin{itemize}[leftmargin=*]
    \item \textbf{Alert and halt cars}  if they don't stop while a wheelchair is crossing the road.
    \item \textbf{Accompany wheelchair users across the road} after pressing a button at the street corner.
    \item Collect crossroad information, such as accidents and \textbf{send the information to the wheelchair users' phone}.
\end{itemize}
&
\begin{itemize}[leftmargin=*]
    \item Indicate the traffic situation using \textbf{both traffic light and alert sound} to accommodate people with visual impairments.
    \item \textbf{Store emergency equipment} which can be accessed by pressing an emergency button.
\end{itemize}
\\
\hline
\end{tabular}
\end{table}

\noindent\textbf{Workshop 4: Snow Plow Robot Needs}

\begin{table}[H]
    \centering
    \begin{tabular}{p{0.48\textwidth}p{0.48\textwidth}}
\hline
\centering\textbf{Functionalities} & \centering\textbf{Interactions} \tabularnewline
\hline
\begin{itemize}[leftmargin=*]
    \item \textbf{Plow snow and spread salt} to make the sidewalk safer in winter.
    \item \textbf{Pull wheelchair users out} when they are stuck in the snow.
    \item \textbf{Operate during low human activity periods} to avoid obstruction due to its large size.
\end{itemize}
&
\begin{itemize}[leftmargin=*]
    \item \textbf{Enhanced visibility in the snow} through standout colors, red lights, and beep sounds (Refer Fig.\ref{co-design:decision-2}).
    \item \textbf{Emergency button} to move the robot and contact operators if it gets stuck or runs out of battery.
\end{itemize}
\\
\hline
\end{tabular}
\Description{This table contains summaries from four workshops focused on identifying user needs and interaction preferences for different types of service robots: 1) Cargo Carrier Robot, which includes functionalities like carrying heavy cargo and adjustable height for accessibility, with interactions such as voice commands and smartwatch control. 2) Grocery Fetcher Robot, designed to pick heavy and high-placed items and integrate with shopping carts, offering natural language understanding and touchpad communication. 3) Crosswalk Guide Robot, aimed at assisting wheelchair users at crosswalks with functionalities like alerting cars and providing traffic information, and interactions including traffic light and alert sounds. 4) Snow Plow Robot, which plows snow and assists wheelchair users in snow, with enhanced visibility features and an emergency button for safety and operational reliability. Each workshop highlights specific functionalities and interactions tailored to improve accessibility and user experience in urban environments.}
\end{table}

\end{figure*}

% They subsequently discussed the feasibility and forms that would be needed for this functionality, with consideration of ways to make it more acceptable to those who might find it frightening. For example, Lily asked Bella ``Should we make it slower to not obstruct others'' and discussed the mechanisms that would be needed to achieve this.

% Arguably, not all the designs created in the workshops were optimal. However, the team's tensions and navigations toward their final decision illustrate the essential criteria a robot company might want to reflect on. These included:
% \begin{itemize}
%     \item ways public robots could be accessible for PwMD
%     \item features that would enhance PwMDs use of public spaces
%     \item features that PwMD would strongly require when interacting with robots in public
%     \item limitations of robotic technologies and possible trade-offs between technical feasibility and PwMD ideal concepts.
% \end{itemize}
% what ways the public robots can bring the value of accessibility to the PwMD in public venues? What are the features that PwMD strongly requires and why? What is the limitation of robotic technology, and how can we trade off, including inventing new? 

\subsection{Summary of the Implications to Future HRI Practice}
\label{co-design: recommendations to the future HRI study}
Abstracting beyond the specific needs of each robot, the collective needs uncovered in the workshops appear to address three areas. First, there is a need for different robots to carry and transport items for PwMD as they are moving about the world. Second, various communication modalities were desired, including voice interaction and touchscreen-based interfaces. Across workshops, it was clear that these should be carefully considered to meet people at their abilities, with examples of smartwatches on the user's wrist or touchscreens that can move to meet the person, rather than requiring someone to reach for a screen.
Thirdly, the workshops revealed a need for robots to clear the sidewalk and make it more accessible for PwMD such as by communicating with other technologies like cars to have them avoid PwMD or by clearing the sidewalk of obstacles.

Overall, these examples suggest opportunities for robots that work on behalf of PwMD rather than simply getting out of the way of people.
While each design concept might suggest specific technical requirements for future public robots, we acknowledge that our results represent only four groups who all ended up having fairly different design ideas.
In lieu of providing specific recommendations, we believe that further co-design sessions among people with disabilities and roboticists can reveal even more opportunities for supportive public robots and that designs addressing the specific use cases and contexts are needed.
Based on our sessions, two areas for future exploration include robots that operate on sidewalks that can communicate with other road agents (as suggested in the crosswalk robot) and robots that can clear debris from a walkway, making it more accessible to others (as suggested by the snowplow robot).
Further, robots aimed at carrying goods might consider how their service could be used for people carrying their own items rather than receiving a delivery.
Finally, all designs suggested clear communication between the robot and people, however, communication modalities should be tested with a diverse set of users and should likely be multimodal to accommodate multiple communication abilities and preferences.
Overall, we believe HRI practitioners leveraging similar co-design methods can uncover new technical requirements for improving accessibility using public robots.

\begin{figure*}
    \centering
    \includegraphics[width=0.65\textwidth]{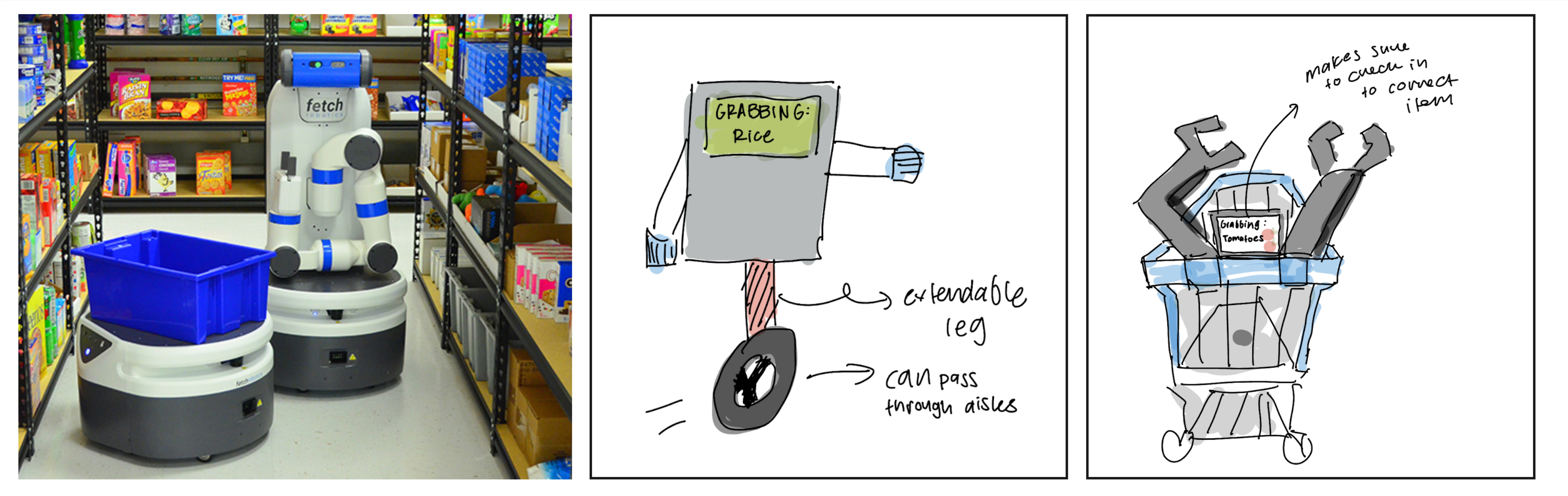}
    \caption{For the grocery fetcher robot, Michael, initially referred to an existing robot product---Fetcher (left image), however, Adrianne argued the base should be narrower to not block ways (middle image). They then decided to utilize the shopping carts as the mobile base (right image). }
    \label{co-design:decision-1}
    \Description{This figure illustrates the co-design process for a grocery fetcher robot, depicted through a sequence of three images arranged horizontally. The first image on the left shows the "Fetcher," an existing robot product initially referred to by Michael. The middle image represents Adrianne's suggestion of designing a robot with a narrower base to prevent it from blocking pathways. The final image on the right exhibits the agreed-upon solution to utilize shopping carts as the mobile base for the robot. The visual development from the initial concept to the final decision showcases the collaborative decision-making process in the design workshop.}
\end{figure*}

\begin{figure}[b!]
     \centering
        \includegraphics[width=0.44\textwidth]{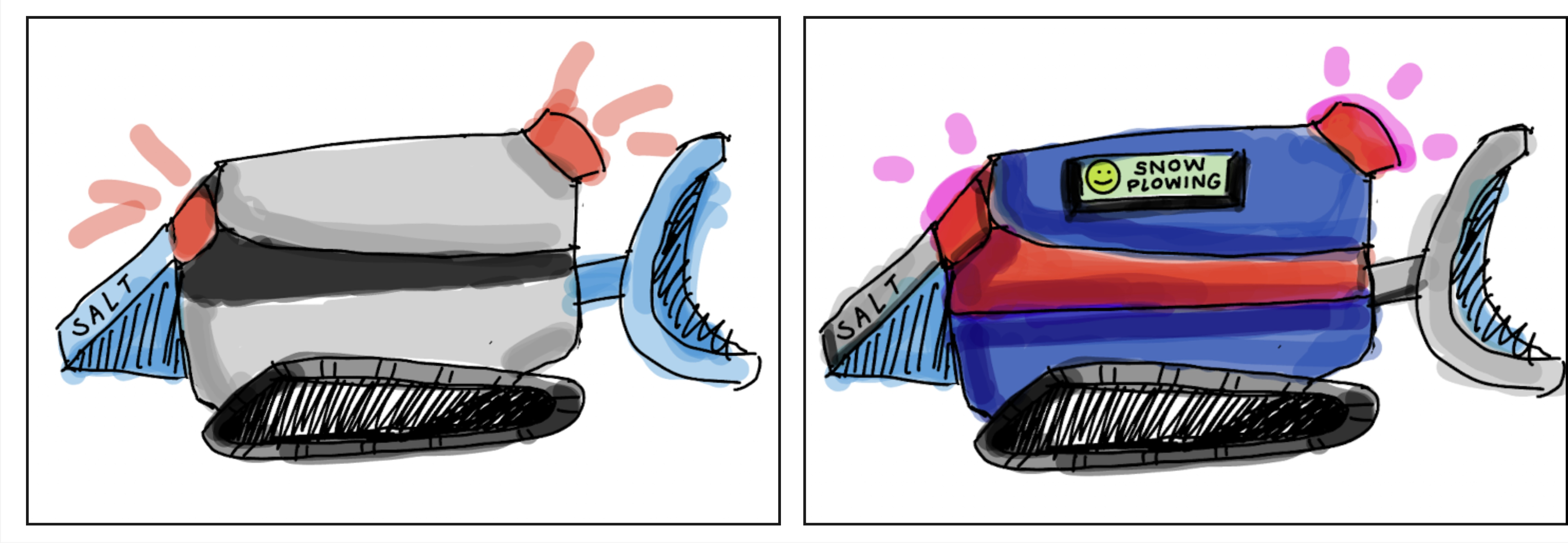}
         \caption{An illustrator sketched the first version of the snow plow robots (left image) based on Will's description. Afterward, Emma and Will found the flaws of robots not being visible in the snow and thus changed the color and proposed adding indicative lights (right image).}
         \Description{The figure contains two images side by side representing different stages of a co-design workshop. The left image displays the first version of a snow plow robot sketch based on a description provided by a participant named Will. The right image shows modifications made to the initial design; the color of the robot has been changed and indicative lights have been added to ensure visibility in the snow. These changes were suggested by Emma and Will to rectify the flaw in the initial design where the robot was not visible in the snow. The changes aim to enhance the robot's visibility in snowy conditions.}
         \label{co-design:decision-2}
\end{figure}

\section{Discussion}
% In the following, we discuss design considerations, the urgency of structured regulations, and ethical considerations pertinent to the deployment of robots in public domains.
Our work explores PwMD's perception of sidewalk robots, revealing inaccessible features and interaction design (RQ1). We also delve into the practice of roboticists, understanding some barriers and opportunities for them to improve accessibility in robot design (RQ2). By bringing PwMD and roboticists together to co-design future public robots, we learn about potential accessibility needs for future robots (RQ3). Below, we discuss the implications of our work and place it in a broader context.
 
\subsection{
Learning from Interviews and Co-Design for More Holistic Public Robots Research Methods}

Inspired by prior qualitative-based HRI studies~\cite{valenciaCodesigningSociallyAssistive2021,bhat2022confused,bennett2021accessibility}, we leveraged semi-structured interviews and co-design workshops to study two sets of stakeholder perspectives concerning emerging public robots, as well as collaborations between them. In our co-design workshop, we intentionally invited robotic practitioners from diverse backgrounds, from design and engineering to industry and academia. We found that their backgrounds and expertise influence their views and, in turn, the co-design outcomes. For instance, a roboticist with extensive experience in proxemics frequently raised questions regarding how wheelchair users navigate robots in public spaces. We interpreted this in two ways: firstly, it shows the opportunity space of incorporating accessibility in different facets of HRI; secondly, future studies should be more mindful of balancing the professional participants to avoid the outcome being over-directed by the participant's personal experience. We recommend future studies adopting iterative and long-term co-design sessions \cite{ostrowski2021long} to facilitate reflective design processes. This can be effective in providing a more comprehensive lens on the accessible feature and also improve the achievability.

Through interviews with PwMD, we learned of potential conflict scenarios between PwMD and the robots (refer to Section~\ref{Conflicts over Sidewalk Resources}) influenced by a range of HRI design factors (refer to Section~\ref{HRI design factor tensions}). Building on these scenarios, future work could concentrate on refining the interaction design of sidewalk robots \cite{jennings2019study}, accommodating the multifaceted requirements of PwMD \cite{carrington2014wearables,li2022freedom}, and taking into account diverse robotic design dimensions such as anthropomorphism~\cite{babaLocalVsAvatar2021} and non-verbal signals~\cite{babelSmallTalkRobot2021,fortunatiMultipleCommunicationRoles2020}
 \cite{li2023understanding,shinohara2011shadow,li2021choose},  \cite{li2022freedom,fan2021eyelid,sun2021teethtap,carrington2014wearables}
Nonetheless, there remains a pressing need to test these interactions and qualitative findings in real-world evaluations. For instance, as discussed in Section~\ref{Usability of different interaction modalities}, our interviews suggested PwMD have a preference for voice interaction, valuing both its inherent naturalness and avoiding direct, tactile engagements while navigating the street. In contrast to our findings, some literature suggests a general expectation that robots remain non-vocal \cite{thunbergArePeopleReady2020}, and that the willingness of individuals to instigate conversations with robots might be minimal \cite{weissRobotsAskingDirections2010,jensenRobotsMeetHumansinteraction2005}. Yet, in scenarios laden with potential conflicts, such as robots obstructing vital sidewalk resources, lacking different forms of interaction, including speech, could lead to issues. Consequently, to truly cultivate an interaction paradigm that is both usable and societally acceptable, it is imperative to delve deeper into the disparities between PwMD's anticipated and actual responses to robots in public via comprehensive contextual evaluations and observations \cite{weinberg2023sharing,gockleyDesigningRobotsLongterm2005}.

\subsection{Inclusive Stakeholder Engagement: A Prerequisite for the Ethical Design of Public Service Robots}
\label{discussion:engagement}
Our interview findings explored opportunities to make public robots more accessible, such as adapting to their mobility aids (Section\ref{Recognizing and adapting to PwMD}), enhancing regulations to move beyond physical size and speed to consider robot behaviors (Section\ref{need of regulation}), and providing mechanisms to address robot malfunctions (Section~\ref{sec:expectation}). However, we need to be more cautious about techno-capitalism: there is a risk that robot companies might leverage these enhancements to assert the accessibility of their robots, which contributes to rationalizing the deployments. Indeed, certain segments of the disabled community have already voiced reservations concerning private-owned delivery robots taking up space on the public sidewalk ~\cite{bennett2021accessibility}.

So, what are the ways that accessible robots can enhance the accessibility of public space? PwMD in the workshops suggested that the robots could enhance their daily living, including improving their carrying capabilities and alleviating transportation anxieties (Section~\ref{co-design:outcome and ideas}). They also collaborated with roboticists, making design decisions that led to accessible features such as adjustable robots and examining the value of the robots within their technical viability (Section~\ref{co-design:uncoveraccessibility}). The accessible features they raised might also challenge current robot capacities and roboticist's conception of universal design (Section\ref{co-design:outcome and ideas}). 
Such co-design engagements may offer a chance for roboticists to learn more about accessibility and reflect on their practices. Thus, to ensure accessibility, it is imperative to engage PwMD both in the evaluation of robot designs and early ideation of robot features and interaction.

We also recognize that solely focusing on people with mobility impairments is insufficient,and may pose a challenge to the generalizability of our findings. For example, in the conflict scenarios discussed in Section \ref{sec:PwMD-concerns}, interviewees highlighted design factors that led to a sense of competition for the curb cut, a resource indispensable to PwMD. However, people with visual impairments may have different perceptions of robot behavior than those seen as obstructive by PwMD. Therefore, design suggestions such as voice interaction may be invalid or even unfavorable to them. Even though the findings presented here are specific to the populations with whom we engaged, our interview design (ref to Section~\ref{methods:PwMD}) is  generalizable as it effectively introduced the concept of public robots, different design factors, scenarios of encountering the robots, and a comprehensive set of interaction modalities. All of our PwMD participants were able to comment on the robot even without prior exposure. Thus, future research can reuse and adapt our interview and co-design methods to engage other groups to compare and extend our findings on accessible public robot design. 

Moreover, findings from our engagements with robot practitioners suggest collaborations with other parties, such as policymakers and urban planners, are also necessary to realizing the vision of accessible robot design (refer to Section~\ref{organizational challenge}). The idea of broader engagement beyond practitioners and everyday people echos recent HRI studies on ethical congruent operations~\cite{ostrowskiEthicsEquityJustice2022,winkleFeministHumanRobotInteraction2023}, which prompts vital inquiries such as "\textit{Who gets to be included in the robot design process?}"~\cite{ostrowskiEthicsEquityJustice2022}. We argue that future robot design should include multi-stakeholder engagements through co-design workshops with appropriately designed participation frameworks~\cite{zytkoParticipatoryDesignAI2022,zhangDeliberatingAIImproving2023} that enable such collaboration.

\subsection{Enhancing Standards, Policies, and Regulations for the Development of Accessible Public Robots}
\label{discussion:guildlines}

Accessibility in robot design and development remains insufficient, even as sidewalk robots are increasingly deployed in public spaces. Prior approaches to ensuring accessibility found in desktop computing accessibility may offer promising opportunities for benchmarking~\cite{vigoBenchmarkingWebAccessibility2013}, scaffolding~\cite{elavskyDataNavigatorAccessibility}, and validating \cite{mankoff2005your} robotic accessibility. Another needed aspect is the development of overarching design frameworks for sidewalk robots that include accessibility recommendations \cite{DesignPrinciplesRobotAssisted}; this is also paramount to nurturing a culture of accessibility amongst roboticists (as described in Section\ref{discussion:engagement}). This will take time and as noted in Section\ref{need of regulation} the lack of overarching design frameworks can be a challenge, especially for start-ups who potentially lack the awareness, bandwidth, and resources to emphasize access in design and development. As such, we saw robotic practitioners advocate for enhanced guidelines and regulations to promote accessibility.

Proactive public policy can do much to ensure that society reaps the greatest benefits from new technology while reducing possible harms ~\cite{cairneyWhyIsnGovernment2020}. Our qualitative data from PwMD reveal that current regulations concerning sidewalk robots may be overly broad and, in some cases, excessively prescriptive. For instance, PwMD's perception of the robot as lacking communication and other necessary interaction functionalities to fully adapt to the complexities of the sidewalk (refer to Section\ref{Conflicts over Sidewalk Resources}) can challenge the sidewalk robot's current classification as a "pedestrian." Some robotic behaviors such as inadvertently blocking access or malfunctions could also violate U.S. mandates that require public sidewalks and services to be universally accessible ~\cite{ADAAccessBuildings}. Thus, beyond simply defining the maximum weight, size, and speeds of the robot \cite{BillI4329789:online}, we argue future public robot regulations should go further and more comprehensively attend to the interactions and effects---intended or unintended---of sidewalk robots to ensure accessibility for all.

Additionally, the synergistic relationship between public robotic policies and urban planning remains under-explored. The rise of public robots could radically reshape the distribution, utilization, and dynamics of urban spaces. We have seen recent smart city practices start to consider establishing a dedicated line on the sidewalk for mobile robots to operate ~\cite{warrenToyotaJustStarted}---our robotic practitioner interviewees also mentioned similar ideas (refer to Section\ref{need of regulation}). However, from an accessibility lens, salient questions arise: Will reallocating public space to robots exacerbate existing accessibility constraints on public sidewalks? To what extent can accessibility challenges be alleviated? And, fundamentally, should urban planning cater to robots, or should robot designs adapt to current urban environments?
% [I will unpack more of the urban space side--- Or maybe not, we are reaching excessively long]

% \section{Future work}

% Additionally, research involving a broader participant sample is crucial to understanding the needs and experiences of various demographic groups. Our research brings opportunities to the HRI and Accessibility communities to ensure that the future of public robots centers on the need for inclusive and equitable pedestrian walkways.
% \section{Limitations}
% We acknowledge the limitations present in our three studies. A primary constraint is the sample sizes, with all participants recruited from the U.S. Our interviews with PwMD involved only 15 interviewees, recruited through social networks, potentially omitting more diverse perspectives. With only four co-design workshops, the results may not be widely generalizable. Future endeavors would gain value from conducting additional sessions and recruiting people with a wider range of disabilities.

\section{Conclusion}
As the use of public robots continues to grow, so does the likelihood that they will encounter people with mobility disabilities. After speaking to PwMD, we discovered that they perceive current sidewalk robot designs as inaccessible. Furthermore, the roboticists that we spoke with suggested that such challenges could only be solved with early and deep participation of PwMD, which current robotic practice may fail to do due to resource constraints and a problematic mindset of patching accessibility only after the issues and harms have unfolded in real-world encounters. By pairing PwMD and roboticists in co-design workshops, we observed how they navigated the development of public robots together in a way where accessibility is centered and the public good is prioritized. Our participants collectively designed accessible features such as fetching groceries and managing car flows. The process also revealed some of the challenges found in such collaboration and in balancing accessibility with other aspects of technical feasibility. Connecting our findings with current robot regulations and ethical design considerations, we believe that solving robot accessibility issues will require involving broader stakeholders when designing a robot and developing better public policy and regulations for robots, robotic practitioners, and the urban space.

\begin{acks}
This work was supported by the National Science Foundation under grant CNS \#2125350 \textit{Smart and Connected Communities: Planning Grant - Equitable new mobility: Community-driven mechanisms for designing and evaluating personal delivery device deployments}. Any opinions, findings, and conclusions or recommendations expressed in this material are those of the author(s) and do not necessarily reflect the views of the National Science Foundation. We acknowledged all the participants for their engagements and insights. We also appreciated the valuable feedback from Henny Admoni and Laura Dabbish. 
\end{acks}

\bibliographystyle{ACM-Reference-Format}
\bibliography{Main}

% \bibliographystyle{ACM-Reference-Format}
% \bibliography{sample-base, Accessibility}
\end{document}